# FORENSIC VIDEO ANALYTIC SOFTWARE

Undergraduate graduation project report submitted in partial fulfillment
of the requirements for the
Degree of Bachelor of Science of Engineering

in

The Department of Electronic & Telecommunication Engineering
University of Moratuwa.

Supervisor:                                        Group Members:

Dr. R. Rodrigo                       R. S. L. D. S. Goonetilleke    130179H
                                     B. Kapilan                     130270E
                                     A. J. Ratnarajah               130514H
                                     M. H. G. D. Tissera            130600T

January 2018

Approval of the Department of Electronic & Telecommunication Engineering

……………………………………………….
Head, Department of Electronic &
Telecommunication Engineering

This is to certify that I/we have read this project and that in my/our opinion it is fully adequate, in scope and quality, as an Undergraduate Graduation Project.

Supervisor: Dr. Ranga Rodrigo

Signature:  …………………………………

Date: ……………………………………….

# Declaration

This declaration is made on January 30, 2018.

**Declaration by Project Group**

We declare that the dissertation entitled *Forensic Video Analytic Software* and the work presented in it are our own. We confirm that:

- this work was done wholly or mainly in candidature for a B.Sc. Engineering degree at this university,
- where any part of this dissertation has previously been submitted for a degree or any other qualification at this university or any other institute, has been clearly stated,
- where we have consulted the published work of others, is always clearly attributed,
- where we have quoted from the work of others, the source is always given. With the exception of such quotations, this dissertation is entirely our own work,
- we have acknowledged all main sources of help,
- parts of this dissertation have been published. (see List of Publications).

………………………..  ……………………………………
Date  R. S. L. D. S. Goonetilleke (130179H)

……………………………………
B. Kapilan (130270E)

……………………………………
A. J. Ratnarajah (130514H)

…………………………………
M. H. G. D. Tissera (130600T)

**Declaration by Supervisor**

I/We have supervised and accepted this dissertation for the submission of the degree.

………………………………….. ………………………………
Dr. Ranga Rodrigo Date

# Abstract

**FORENSIC VIDEO ANALYTIC SOFTWARE**

Group Members: R. S. L. D. S. Goonetilleke, B. Kapilan, A. J. Ratnarajah, M. H. G. D. Tissera

Supervisor: Dr. Ranga Rodrigo

Department of Electronic and Telecommunication Engineering

Law enforcement officials heavily depend on Forensic Video Analytic (FVA) Software in their evidence extraction process. However present-day FVA software are complex, time consuming, equipment dependent and expensive. Developing countries struggle to gain access to this gateway to a secure haven. The term forensic pertains the application of scientific methods to the investigation of crime through post-processing, whereas surveillance is the close monitoring of real-time feeds.

The principle objective of this Final Year Project was to develop an efficient and effective FVA Software, addressing the shortcomings through a stringent and systematic review of scholarly research papers, online databases and legal documentation. The scope spans multiple object detection, multiple object tracking, anomaly detection, activity recognition, tampering detection, general and specific image enhancement and video synopsis.

Methods employed include many machine learning techniques, GPU acceleration and efficient, integrated architecture development both for real-time and postprocessing. For this CNN, GMM, multithreading and OpenCV C++ coding were used. The implications of the proposed methodology would rapidly speed up the FVA process especially through the novel video synopsis research arena. This project has resulted in three research outcomes with pending conference paper publications titled Moving Object Based Collision Free Video Synopsis, Forensic and Surveillance Analytic Tool Architecture and Tampering Detection Inter-Frame Forgery.

The results include forensic and surveillance panel outcomes with emphasis on video synopsis and Sri Lankan context. Principal conclusions include the optimization and efficient algorithm integration to overcome limitations in processing power, memory and compromise between real-time performance and accuracy. This report discusses the progress of the Final Year Project, Forensic Video Analytic Software.

Dedicated to

Our lecturers, parents, siblings and friends

# Acknowledgments

We express our sincere gratitude to Dr. Ranga Rodrigo, our project supervisor, for his immense guidance, support and motivation throughout the project.

A special thank you goes out to our department head, Prof. Rohan Munasinghe for providing laboratory facilities since the commencement of the project.

Our gratitude is also extended to Dr. Pujitha Silva, Dr. Peshala Jayasekara and Dr. Nuwan Dayananda for their valuable and constructive insight during our Feasibility, Progress-review and final presentations.

We extend our thank you to Dr. Anjula De Silva, Coordinator of EN 4202: Project module for his guidance and instruction.

Finally, we convey our heartiest gratitude to our parents, siblings, batchmates and friends for their loving support and encouragement.

# Table of Contents

# List of Figures

# List of Tables

# Acronyms and Abbreviations

BOF – Bag of Feature

BRISK – Binary Robust Invariant Scalable Key point

CCTV – Closed-Circuit Television

CNN – Convolutional Neural Networks

FREAK – Fast Retina Key point

GMM – Gaussian Mixture Model

GPU – Graphics Processing Unit

GUI – Graphical User Interface

HoG – Histogram of Gradients

MIL – Multiple Instance Learning

SORT – Simple Online and Real-time Tracker

SVM – Support Vector Machine

YOLO – You Look Only Once

# Chapter 1

## INTRODUCTION

This chapter commences with an overview of the forensic ideology with background information. Then the problem statement, primary objectives and scope of the project are presented. Next a review of the pertinent literature, followed by the used methods of investigation. The chapter ends with the principal results of the investigation and conclusion.

### 1.1 What is Forensic and Surveillance?

Forensic denotes the application of scientific methods and techniques to the investigation of crime through post-processing. Surveillance relates to the close monitoring of real-time video feeds. The key difference between forensic and surveillance lies in the analysis stage. Forensic pertains to post processing whereas surveillance is in real-time.

### 1.2 The Problem Statement

There is a necessity for law enforcement agencies such as Sri Lankan Police and security officers to have an integrated and efficient Forensic Video Analytic (FVA) software to process CCTV video feeds both in real-time and for post processing. This is required for evidence extraction and prevention of criminal actions.

The following are persistent problem in the FVA scenario.

- Currently there are millions of CCTV cameras in Sri Lanka which capture data continuously for 24 hours throughout the year. However, there is a scarcity of human resources to watch the entire duration of videos and detect minor or major forensic activities. Since human attention would be lost after 20 minutes, most of the recorded videos are considered as insignificant. Our implementation of a novel technique called video synopsis, can be used to summarize a 24-hour video into a short video clip of few minutes, while preserving all human activities so that security and police officials can watch CCTV videos within a short duration and detect any forensic activities, thus saving time, effort, money and speeding up capture and conviction of criminals.

- Abnormal activities usually would precede criminal actions. Hence detecting anomalies well in advance would assist in crime mitigation by sounding an alarm. Further self-learning in a particular user-defined environment to detect any anomaly behavior is not in the market at present.

- Criminals may produce false evidence by tampering the original CCTV coverage when presenting to the jury. Hence officers need to verify whether the produced evidence has been tampered or not. Also, false accusations could be discarded quickly.

- Inability to recognize vehicle number plates and suspicious faces in CCTV video due to degraded quality of the CCTV output can be prevented. Therefore, we are proposing image enhancement techniques, which can enhance important regions of frames and hence would be helpful for police officers to collect data, such as the face of the culprit.

- Unavailability of software to assess the reliability of evidences produced in court as images and videos.

- Lack of security tools that can be concurrently used for surveillance as well as forensic analysis. Further usage of multiple prominent features is limited and now a days several independent software would be required to post process.

"Developing an integrated and efficient forensic video analytic software to analyze CCTV feeds both in real-time and for post processing." would solve such dire encounters.

**1.3 Primary Objectives**

Our primary objectives intend wide coverage of forensic and surveillance analysis applications, while integrating multiple forensic and surveillance techniques in an efficient manner. The key criteria identified could be categorized into multiple object detection, multiple object tracking, anomaly detection, activity recognition, general and specific image enhancement, tampering detection and video synopsis. These procedures require high processing capability and were addressed through optimization techniques and GPU implementation of common and frequent processing steps.

Our final output is a deployable software efficiently integrating all the above-listed functionalities, that could operate on workstations with a user-friendly GUI connected to a CCTV camera. Along with this our deliverables achieved include GPU implementation of common and frequent processing steps with optimization of machine learning techniques as well as publications in terms of three conference papers.

**1.4 Scope**

Although our project title is very general in nature and spans a broad scope, it was limited by considering the significance and importance of multiple forensic and surveillance applications. Our scope can be categorized into two, namely Forensic Detection and Surveillance.

*1.4.1 Forensic Detection*

Under the forensic detection scope, CCTV videos would be post-processed to detect any crime scene and to collect digital evidence. The forensic detection approach falls into the following main categories.

- Video synopsis: This is a novel approach of summarizing a long video into a short video clip.

- General image enhancement: A frame or image or even a cropped-out region of interest is enhanced to extract more information with user input as guidance.

- Face enhancement: The face of a person in an image or in several frames of a video is enhanced here.

- Textual enhancement: This is an approach of correcting the textual area of an image which was degraded by motion blur and gaussian noise.

- Tampering detection: This detects whether a given video, or an image is real or edited. In a given video, frame segments could have been inserted, deleted or repeated. In an image or particular frame of a video, same area could have been copied and pasted to another region or some areas could have been covered by regions from different images.

*1.4.2 Surveillance*

The surveillance activities include real-time object detection and tracking of multiple objects and activity recognition.

- Anomaly detection: Prohibited motions or strange movements are detected using optical flow techniques. Such detected motions are tracked throughout the rest of the video.

- Trespass detection: It is a manual user interface where, motion in user defined boundaries are alarmed. It can be used in railway crossings.

- Object detection: This involves detecting bounding boxes of objects appearing in frames and the recognition of those objects. It is used as a supplementary block for object tracking and selective image enhancement.

- Object tracking: Any objects detected or selected by the user is tracked along multiple frames of the video feed continuously in real-time.

- Camera tampering detection: Camera redirection, camera shaking and camera blocking are detected and alarmed in real time.

- Activity recognition: Human interactive behaviors such as fighting, robbing and meeting as crowd and individual behaviors like waving hand and clapping are recognized.

**1.5 Review of Pertinent Literature**

In our project we have used moving object-based video synopsis approach mentioned in [1]. Background subtraction based on Gaussian Mixture Model was used to detect moving objects and generate a proposed tracking algorithm to track them in subsequent frames. Since [1] creates synopsis video once all the objects have been tracked, it is inefficient. An online based video synopsis approach has been adapted, where tube generation through tracking and tube re-arrangement occurs in parallel.

Existing approaches have been modified and video synopsis algorithms proposal based on selecting cluster of tubes and re-arranging them is underway. Our proposed method was tested with dataset in [2]. We were able see that our algorithm was able to work in real-time and it can efficiently reduce spatial and temporal redundancies than [2] and [3].

**1.6 Used Methods of Investigation**

Under surveillance, objects detected with You Look Only Once (YOLO) [9] are tracked using Simple Online and Real-time Tracker (SORT) [14]. Here Kalman filter for motion estimation and Hungarian algorithm for data association is implemented. Activity recognition is done by feature extraction (BRISK and FREAK) [15] and bag of feature representation followed by SVM based classification. For anomaly detection optical flow based clustering is used.

For video synopsis, since collisions are visually displeasing and energy minimization approach is time consuming, our project uses a less complex approach which can create collision free summarized video clips based on moving objects.

Adjustments in contrast, exposure, Hue, Saturation, Value (HSV), temperature, noise reduction, shadow/highlight and sharpening are provided with user preference sliders for general image enhancement. Face enhancement was undertaken using linear interpolation followed by super-resolution. Textual enhancement used Wiener filter algorithm to efficiently remove motion blur and gaussian noise simultaneously. For this image was deconvolved using point spread function signal to noise power ratios estimated by the user.

Tampering detection covers three areas, namely inter-frame forgery detection, intra-frame forgery detection and camera tampering detection. To detect inter frame forgery a dense optical flow based technique is used where optical flows for each frame is calculated and the optical flow variance sequence is analyzed for discontinuities. Under intra-frame tampering copy-move forgery is detected with a key point SIFT [32] based approach. Splicing detection involves analyzing DCT histograms [34] to detect double quantization effect of JPEG compression. Camera tampering is detected by thresholding the foreground extracted contour area of moving objects.

## 1.7 Principal Results of Investigation

Experiments were conducted on two datasets containing a total of 6 videos from existing research papers on video synopsis. The resulting video synopsis clips were more than 4 times shorter and visually pleasing.

The algorithms have been tested in Sri Lankan context for different faces, number plates and acceptable results have been obtained. Under surveillance all the included modules were tested with Sri Lankan context and acceptable results were obtained.

## 1.8 Summary

Forensic Video Analysis is a growing security concern of society at large. The expected deliverables of the project include a complete deployable software with a user-friendly GUI containing optimized functionalities categorized under the scope of the project and a conference paper. This chapter provided an introduction to the overall project with insight into methods, sub-blocks and results.

# Chapter 2

## LITERATURE REVIEW

The field of forensic analytics and surveillance is a vast area with a lot of implementations and with a variety of techniques spanning a long history. Our focus is on an integrated and efficient software which combines these different techniques together. We categorize our scope into two main streams, namely forensic and surveillance. Forensic analytics involves post processing of video feeds while surveillance techniques require the video to be processed in real time. The focus of this project is on seven techniques which are related to forensic and surveillance aspects of CCTV video processing. They are video synopsis where hours long videos are summarized to several minute videos without losing important information, object recognition where the objects on the video are classified into categories, object tracking where the detected objects are tracked along their movements, activity recognition where the activities of people involved are detected, anomaly detection where abnormal behaviors of people are detected based on their activities, image enhancement where the images are processed to retrieve hidden details and finally tampering detection where videos and images are checked for manipulation. This chapter covers the pertinent literature review of each stage as categorized under the project scope.

**2.1 Video Synopsis**

Although billions of surveillance cameras are installed spanning many roads and buildings, most of the recorded videos are shelved without being watched or examined. Therefore, most of the criminal activities remain secretive, unless someone complains about a forensic activity. From early days there were many approaches, such as fast forwarding [4] and Optical flow-based motion analysis to select the key frames [5]. The above approaches focus on the temporal reduction of the video.

An approach to video synopsis which optimally reduces the spatio-temporal redundancy in video is presented in [1]. In the above approach video synopsis is itself a video expressing the dynamic movement of the scene. Spatio-temporal redundancies are reduced as much as possible using Energy Minimization approach. Since conventional tube rearrangement based on minimizing global energy function is computationally intensive and time consuming, video synopsis based on potential collision graph was proposed in [6].

There are two approaches of creating video synopsis [7], namely are online and offline methods. Offline methods [1] have two phases of processing. In the first phase the video is scanned through in advance and both trajectories and background are captured and stored. In the second phase all the object tubes are rearranged together by minimizing a cost function at once. Since the above method requires hefty calculations and large memory when processing a long video, online video synopsis method is preferred instead. In this method both phases would be processed in parallel.

**2.2 Multiple Object Recognition**

Our literature review revealed that the two most outperforming candidates for object detection are YOLO [9] and Region based Convolutional Neural Networks (RCNN) [10].

The two current evolutions of YOLO are YoloV2 and Yolo9000 [11]. In this technique of object detection, the image is first fed to a single neural network where the image is divided into regions with bounding boxes and associated probabilities. Predicting the bounding boxes and assigning the class probabilities are done simultaneously making YOLO one of the fastest algorithms for object detection.

The two latest evolutions of RCNN are Faster RCNN [12] and Mask RCNN [13]. In RCNN framework takes an input image, extracts region proposals and compute features for each region using a convolutional neural network and finally classifies each region using class specific linear support vector machines.

Comparing the two competing techniques, the faster algorithm is YOLO while RCNN can be trained to have more accurate results. In real time processing of CCTV feeds, time for processing is a crucial step while in forensic post processing accuracy counts.

## 2.3 Multiple Object Tracking

Multiple Object Tracking is a technique used to interpret how one's visual system tracks several moving objects. This was intended as a real-time surveillance implementation. Online tracking refers to the usage of only current and past frames for tracking. [14] denotes Simple Online Real-time Tracker (SORT) algorithm for a fast and computationally less complex technique to update trajectory estimates. It is a tracking-by-detection method and is dependent on detection quality.

Additionally, there were many trackers such as Kanade-Lucas-Tomasi (KLT) and Mean shift trackers that were tested and eliminated based on performance, speed, memory and accuracy constraints.

## 2.4 Activity Recognition

Activity recognition is recognizing certain human behaviors based on temporal and spatial analysis of a given video output and then comparing them to trained datasets. Our basic approach is to extract dense trajectories-based action tubes from training datasets and extracting BRISK features and FREAK extractors [15] and using them to build bag of feature [16] classification of video using clustering techniques. Then the bag of feature is used to train the SVM classifier which is in turn used in classification of test videos. There are many open source datasets available for activity recognition [17]. However, the dataset we are using must be similar to that of CCTV video feed. Therefore, we selected BEHAVE [18] and CAVIAR [19] datasets to train our network. Although HOLLYWOOD dataset provides plenty of training samples they cannot be used as they are extracted from selected Hollywood movies hence they are non-realistic.

## 2.5 Anomaly Detection

Anomaly detection is the process of detecting some anomalies in the videos like disruptive human behaviors and vehicle motions. E.g. Vehicle moving in wrong direction along a one unidirectional road and peoples exiting form wrong directions. The procedure follows a training part where the optical flow histograms of the videos are calculated block by block and maximum of them are recorded in a busyness matrix and during the testing the if the optical flow histograms are outlying the busyness matrix the anomaly alarm is raised [20]. Datasets for anomaly detection are available for free. We use UCSD anomaly detection dataset [21].

*2.5.1 Trespass Detection*

Detecting the movements in some prohibited areas like railway gates and museum restricted areas. Here boundaries for trespass detection are given by the used though UI and foreground is extracted in those areas and the motion in those areas are detected and alarmed.

**2.6 Image Enhancement**

Under forensic analytics, once the user selects a frame or a region of interest to be enhanced, adjustments in exposure, contrast, saturation, noise reduction, highlight and shadow recovery, sharpening, color temperature, etc. would be facilitated.

*2.6.1 Face Enhancement*

Faces of CCTV footage can be incomprehensible which makes criminal identification cumbersome. [22] identifies a method of enhancement using linear interpolation, followed by super-resolution. Super-resolution is done for several frames of a video or for an image replicated with slight random degradations.

*2.6.2 Textual Enhancement*

According to [23] when capturing the vehicle images by CCTV camera, there are two types of image grading such as motion blur and out of focus blur. [24] describes on using wiener filter for motion blur removal in a noisy image using estimated noise to signal ratio. [23] also describes a general method of out of focus blur removal which is based on unsharp masking. Here, we subtract smoothed version of image from the original image and then we multiply the resulted image by a scalar constant and we add it to the original image.

**2.7 Tampering Detection**

Tampering detection is a set of techniques which are used to identify manipulations of videos done in diverse ways. Under tampering detection, we focus on inter frame and regional tampering as forensic techniques and camera tampering which is a surveillance technique.

*2.7.1 Inter Frame Tampering*

Inter frame tampering is manipulation of frame sequence in a video. Frames can be inserted, deleted, duplicated or shuffled. To address these scenarios, techniques as per [25] such as optical flow, correlation between suspicious frames and consistency of velocity field intensity can be used. Numerous techniques exist in the literature to address the detection of inter frame forgery. [26], [27] and [28] uses optical flows to detect inter frame forgery where there will be a discontinuity in the optical flow variance sequence where the frames have been replaced, deleted or inserted. [29] detects inter frame forgery by analyzing velocity filed consistencies. [30] checks double MPEG compression to detect frame manipulations.

*2.7.2 Regional Tampering*

Under regional tampering we focus on copy paste forgery and splicing. Copy paste forgery means copying a portion of frame and pasting in another part of the same frame or pasting in another sequence of frames. Splicing is similar except that the portions are pasted from a different video. Copy move forgery has been detected by several techniques. These include block matching techniques such as [31] and key point matching techniques such as [32] where they use SIFT features to match copied regions. Splicing is also addressed via different approaches. [33] reveals a comparison between existing techniques which use noise patterns, image compression effects, color filter arrays etc. All these techniques are based on the assumption that the spliced area will be different in terms of a fundamental aspect. [34] uses DCT coefficients to detect double quantization effect of JPEG compression. [35] detects disturbances in the image CFA interpolation patterns by using gaussian distributions where [36] models the local image variance in order to detect image splicing.

*2.7.3 Camera Tampering*

Camera tampering falls under surveillance in which real time video processing takes place and alarms the user. Camera tampering includes camera shaking, camera redirection, camera block and camera defocus. [37] proposes a method to detect camera tampering by robust comparisons of recent and older frames of video using three different measures of image dissimilarity. [38] proposes adaptive techniques to detect camera occlusion, defocus and displacement.

**2.8 System Architecture**

There have been several implementations of forensic and surveillance architectures. However, according to our literature review none of the existing systems cover both forensic and surveillance aspects. Most of the systems were designed to cater to particular needs and were focusing on a small part of either forensic or surveillance areas. Multiscale spatiotemporal tracking through the use of real-time video analysis, active cameras, multiple object models, and long-term pattern analysis to provide comprehensive situation awareness is discussed in [46]. IBM smart surveillance system (S3) [47] provides the capability to automatically monitor a scene and to manage the surveillance data, perform event-based retrieval, receive real time event alerts through standard web infrastructure and extract long term statistical patterns of activity.

**2.9 Summary**

Our literature review basically divided the scope of the project into two main categories, namely forensic and surveillance. Under forensic aspect, this chapter reviews techniques on tampering detection, image enhancement and video synopsis. For surveillance, methods on multiple object detection, multiple object tracking, activity recognition and anomaly detection are reviewed. These seven techniques were critically analyzed in our literature review in order to identify the most efficient algorithms and alternatives. Finally, a review of existing systems integrating forensic and surveillance techniques were reviewed.

# Chapter 3

## METHODOLOGY

This chapter represents the methodologies used in the project under Forensic and Surveillance criteria. The first part of this chapter provides the approach followed. Then forensic and surveillance architectures are separately analyzed. Next the procedures followed under each subtopic are discussed with an in-depth analysis of the proposed video synopsis research area. The latter part of the chapter includes GPU specificities, development tools and conclusion.

**3.1 Approach**

Our project analyses CCTV footage and processes in real time for surveillance applications. Any abnormal behavior would be identified by anomaly detection and such objects would be detected and tracked along subsequent frames. Further the user can select any suspicious object to be tracked continuously. Under trespass detection, the user can draw any region of interest and if any movement is detected in that region, an alarm would be raised. Camera tampering for redirection, shaking or covering the camera would also be detected and alarmed. Further the user can hover the mouse pointer over the frame to observe a zoomed and enhanced region separately. The video feed could be saved for the post processing.

Under forensic analysis, post-processing of videos and images are accounted for. The received video footage would be motion detected and the frames with motion will be permitted for further processing, meanwhile the complimentary frames will be used to model the background. Next these frames will be processed to detect and track objects and create a video synopsis. A frame of a video can be selected for enhancement. Further a region could be cropped out for enhancement. Specific image enhancement categorized as face enhancement and textual enhancement are adaptable based on user input. In addition to enhancement, images could be verified for intra-frame tampering detection. Inter-frame tampering detection is applicable to videos.

Real time or surveillance application includes multiple object tracking, anomaly detection, activity recognition, tampering detection and video synopsis, meanwhile post processes include general image enhancement, face and textual enhancement, multi object detection, tampering detection and video synopsis.

## 3.2 System Architecture

The overall system architecture is two-fold based on the analysis purpose. Forensic and surveillance architecture are interrelated, with real-time footage being stored for future processing as forensic evidence.

### 3.2.1 Forensic Architecture

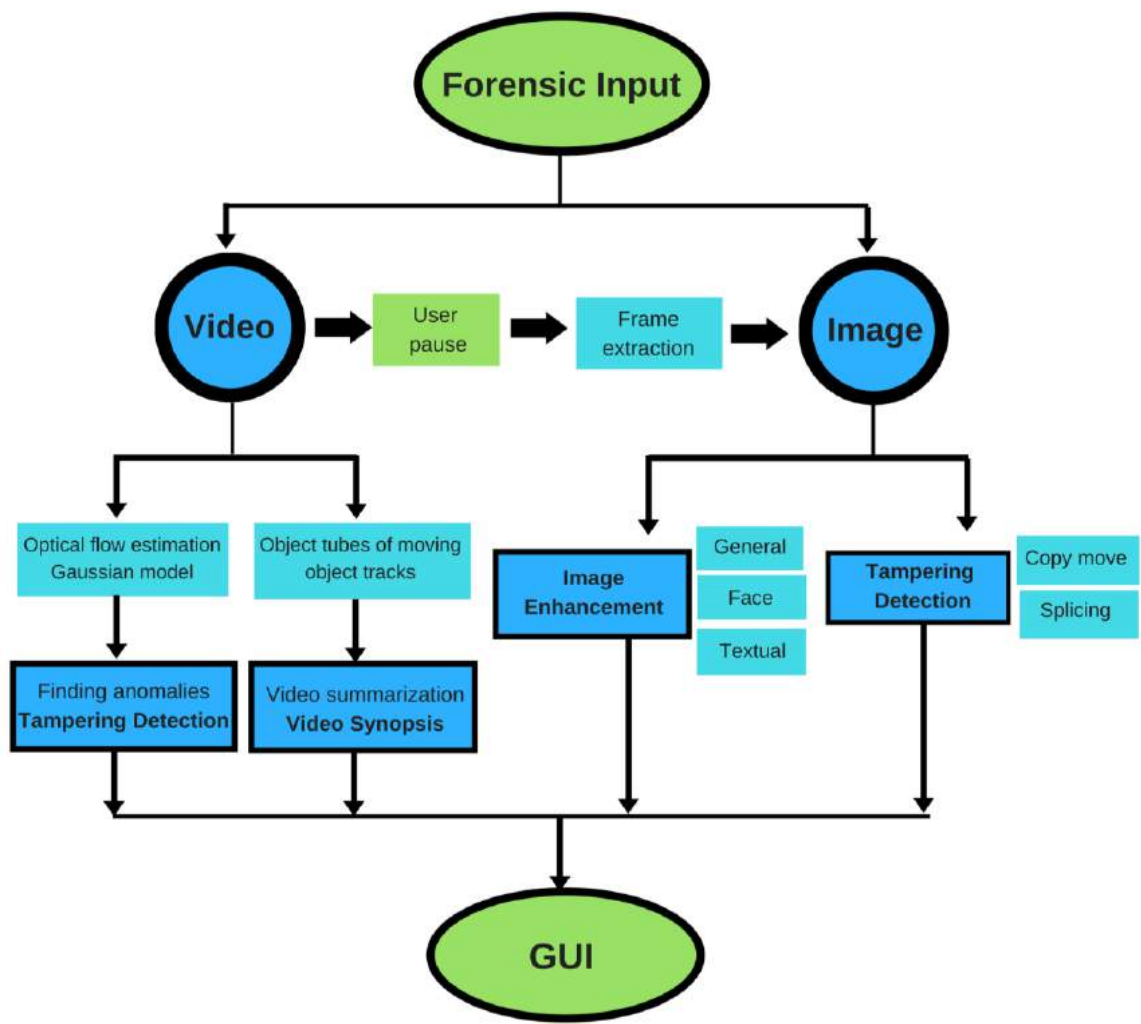

Fig.3.1 Forensic system architecture

We have implemented a unique forensic architecture where the user can post process entire video, frame from a video or any image to collect forensic evidence. In our software the user can summarize a long video into a short video clip consisting of all dynamic movement. Then the user can save time by running the summarized video to observe any abnormal behaviour. When the user sees any suspicious dynamic behaviours, the user can click the object of interest in the summarized video to obtain the complete trajectory of the object of interest in the original video. When objects cannot be recognized properly, the user can pause at the frame of interest and enhance that frame using general enhancement techniques. The general enhancement tool permits changes in contrast, exposure, temperature, sharpness, hue, saturation and value of the image to enhance the image. It also allows recovery of shadows and highlights and reduction of noise.

When the user wants to see the face or textual information in an image clearly, the user can crop the region where the face or textual information is present and enhance it using special enhancement tools. They are made to enhance textual information using Wiener filter algorithm and enhances faces using super resolution algorithm.

The forensic tool can be used to check for image and video tampering as well. The tampering activities detected under the forensic block mainly involve post processing. Tampering detection which falls under surveillance is discussed below under surveillance architecture. In this case, the user can input a video to check whether the video has been tampered frame wise and this includes frame insertion, deletion and repetition. Additionally, the user can input an image to check for image tampering such as copy move forgery, where one region can be copied and pasted covering another region in the same image and splicing where regions of an image can be replaced with portions of other images. These two forgeries can also be used for videos frame by frame where the user must pause the video and check for image tampering in a frame of choice.

*3.2.2 Surveillance Architecture*

In surveillance due to real-time functionality, speed is of the essence. Even a slight delay would cause a backlog in serving the purpose. Hence several parameters such as computational complexity, accuracy, speed, processing power and memory are compensated to obtain the optimum mix. This pertains to real-time videos with continuous monitoring. As per the norm, surveillance commences with multiple object detection. Here YOLOv2 is used for pre-trained object detection.

Thereafter based on the detection probability, and objects recognized, multiple object tracking commences. The tracking ID is assigned using Simple Online Real-Time (SORT) tracker with the aid of Kalman filtering and the Hungarian algorithm. In a parallel thread camera tampering detection is performed. If the area of motion regions of successive frames exceeds a threshold value, it is ascertained that the camera under consideration has been externally tampered with and thus an alarm is raised.

Anomaly detection pertains to abnormal motion is unauthorized directions. Correct movement are initially trained. When any undue activity arises, an alarm is raised followed by the tracking of the anomaly. Under trespass detection the user initially draws the region of interest which could be a traffic offence, a restricted area in a museum, etc. If any movement is detected an alar is raised. At any instance the user can select any object or frame for postprocessing as indicated above under forensic analysis.

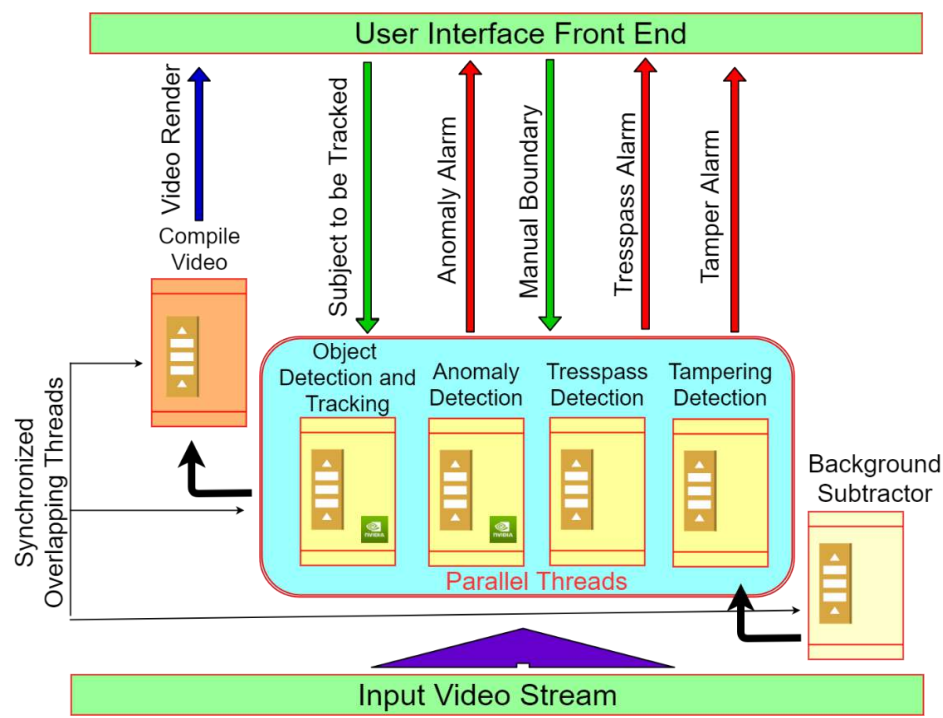

Fig.3.2 Surveillance system architecture

In Fig.3.2, the background subtractor module is needed by all the parallel threads except object detection and tracking. Hence background subtraction is done first through a synchronized thread and then the 4 modules commence. Multiple object tracking with detection, anomaly detection, trespass detection and tampering detection are parallelly executed in the second level thread. Meanwhile the user interface (UI) undertakes communication between these threads as well. Final thread level combines these results and video output is rendered in the Graphical User Interface (GUI). GPU acceleration is utilized for anomaly detection and object tracking to speed up the process.

Fig.3.3 below describes the structure of the GUI designed for surveillance panel.

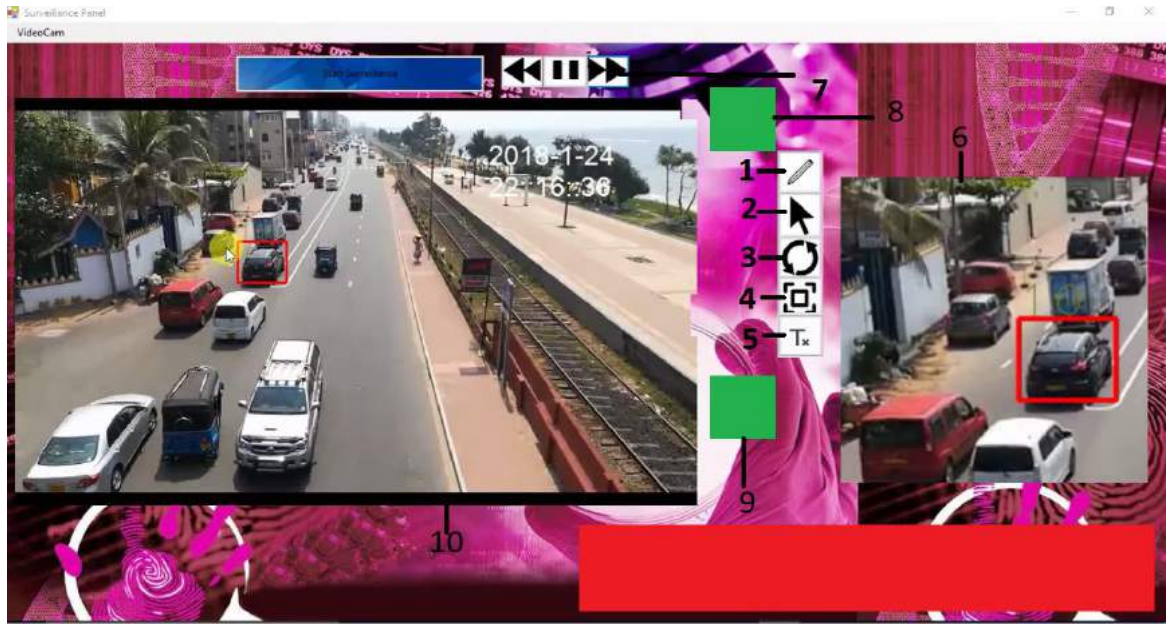

Fig.3.3 Surveillance panel GUI

1- Manual boundary drawing tool for Trespass detection

2- Pointer for on-demand object tracking

3- Refreshing all the settings to default

4- ROI enhancement

5- Switch to original surveillance video

6- ROI window

7- Navigation panel

8- Trespass alarm

9- Tamper alarm

10- Main surveillance window

This GUI is expandable for added tools and features. Designed for single stationary camera surveillance, it is feasible to extend this for multi camera scenarios in the future.

**3.3 Video Synopsis**

*3.3.1 Overview*

The aim of video synopsis is to summarize a long video into a short video clip by reducing spatial and temporal redundancies. The objects are represented as tubes in space-time volume. In our proposed approach the tubes are rearranged such that the time occupied by all the tubes are minimized, whereas the space occupied at each time instants are maximized, without changing the actual spatial location of each tube.

In the first phase of our proposed approach foreground is extracted from the background using Gaussian Mixture Model. Then the binary image generated from foreground and background is dilated and eroded to remove noise and to isolate the individual elements. Contours are generated from the above processed image by joining all continuous points along the boundary. Then convex hulls are generated from the contours and each object is extracted as smallest rectangular boxes enclosing each convex hull. To create tubes for each object, the same objects are tracked throughout the frame. Each tube is given an identification number based on their chronological order.

In the second phase, from the cluster of tubes, each object frame in each tube is selected in First in, First out basis. Maximum number of selected object frames are placed in a background frame with no collision. During placement, higher priority is given to object frames from the tubes that were detected earlier. The proposed approach maintains constant number of tubes in a cluster. Once an object tube in a cluster is completely placed, a new tube will be introduced to the cluster based on their chronological order, thus maintaining the constant cluster size.

*3.3.2 Moving Object Detection*

Since the primary purpose of moving object detection is to detect any moving objects in a frame and object labels are not required, background subtraction method serves this purpose. Background subtraction can be carried out at pixel level or at region level. [39] describes region level approach where region of interest is divided into small blocks of interest and background is modelled using the intensity variance of the blocks. Although this approach is low in computational complexity, it fails to separately detect multiple objects which are nearby, with a distinct bounding box.

To avoid detection of several objects as a single object, pixel-based approach has been adapted in this work. Mixture of gaussian based approach implemented in [40] is used in our approach. Pixel based approach mentioned in [41] and [42] have been used to segment background and foreground based on gaussian mixture. In [40] the number of mixture components have been determined dynamically, per pixel. In this methodology last 100 samples are used as training data and the density is re-estimated whenever a new sample replaces the oldest sample in the training data. Optimum threshold for the squared Mahalanobis distance is set to 25 for the experimented video datasets. In the above implementation shadow is detected if the pixel is darker than the background. Here the shadow threshold of 0.5 is set implying that if a pixel is more than twice darker, it is not considered as a shadow.

Once the foreground and background has been segmented, a binary image is created by assigning one for foreground pixels and zero for background pixels. Then the binary image is dilated and eroded to close holes in foreground image and to remove noise. Then contours have been generated by joining continuous points along the boundary of each foreground objects. Using the contours, convex hulls are generated for each foreground object and each moving object is segmented as smallest rectangular boxes enclosing the convex hull to create tubes.

*3.3.3 Multiple Object Tracking*

Multiple object tracking consists of mainly 2 parts. They are object detection and motion prediction. [43] presents an approach where objects are recognized by comparing an analytical approximation of the skeleton function extracted from the analyzed image with that obtained from model objects stored into a database and then tracked using extended Kalman Filter approach. The dataset used by us to compare the tracking accuracy [44] also uses Extended Kalman filter approach to track.

We present a less complex motion prediction approach which performs well in highway vehicle tracking and at adequate accuracy when objects move in nonlinear motion. In this approach we consider the center of the bounding box of last tracked object in last 10 frames and we predicts its new position.

Let,

i - current frame number

C[i] – center of the tracked object in $i^{th}$ frame

P[i] – predicted center of the tracked object

D[i] – difference between predicted center and current center

If i ≤10

$$D[i] = \frac{\sum_{n=0}^{i-1}(C[n+1]-C[n])*(n+1)}{\sum_{n=0}^{i-1}(n+1)} \quad (3.1)$$

Else

$$D[i] = \sum_{n=1}^{10} \frac{(C[i]-C[i-n])*(10-n)}{45} \quad (3.2)$$

$$P[i] = C[i] + D[i] \quad (3.3)$$

After predicting the future center position of currently tracked objects, this approach maps them with the future detected objects. As one of the characteristics of tracking is to remove noisy detection, this approach gives an Object_ID to detected tubes only if the objects has been consecutively tracked at least

"N" number of frames. Here "N" is calculated as follows,

$$N = framerate * 0.5 \qquad (3.4)$$

Fig.3.4 depicts the results of the multiple object detection and tracking approach proposed us.

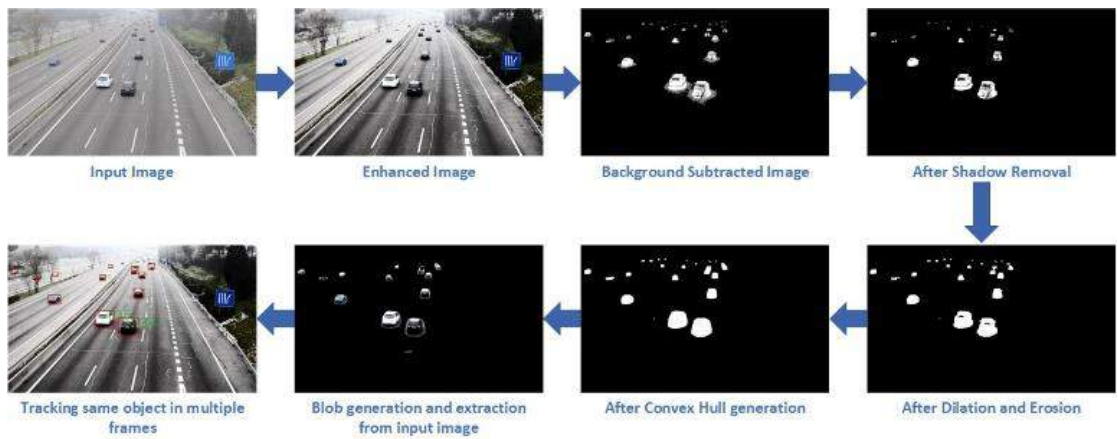

Fig.3.4 Multiple object detection and tracking

*3.3.4 Object Based Video Synopsis*

In the online approach of object-based video synopsis both tube generation and tube re arrangement happens in parallel.

While multiple objects are being tracked in each frame the object tubes are being generated and background image is being stored. In this approach tubes are being re-arranged while the complete new tubes are being generated in parallel.

Here the user defines size of the cluster of tubes to be processed at any given time. Maximum number of objects placed in a synopsis video is positively correlated while the speed of synopsis video generation is negatively correlated with the cluster size. Synopsis video is being created by placing maximum number of objects frames as possible in each frame in their sequential order in the tube with no collision. In this approach the tubes are being placed synopsis video in First in First out basis.

Fig.3.5 shows a graphical representation of video synopsis created using tubes generated from M30-HD video [44].

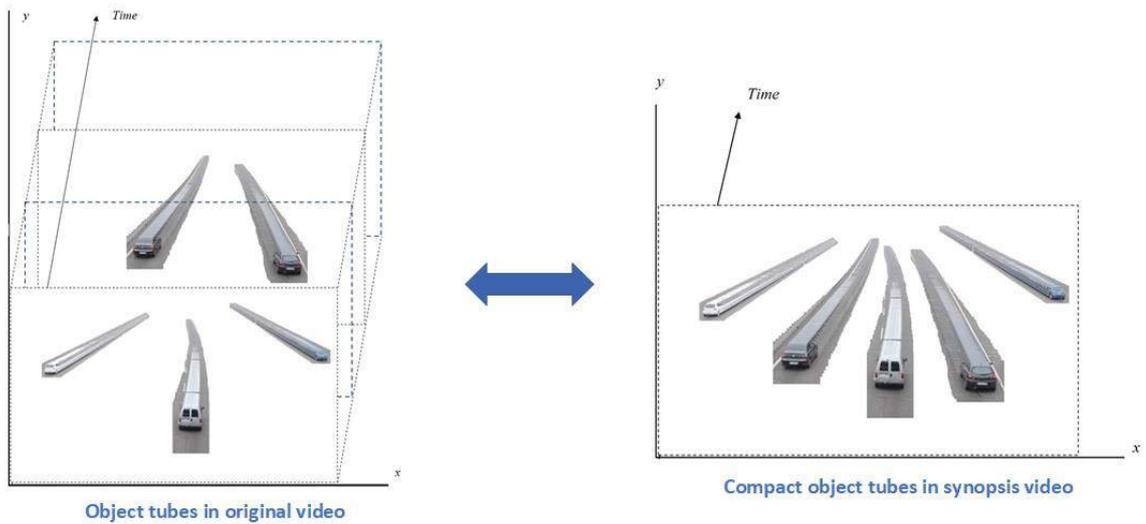

Fig.3.5 Synopsis video generation from object tubes

Algorithm of Track re-arrangement

While (tube generation is not completed, and all tubes are not completely being placed in synopsis video)

{

1. Select a cluster of tubes not completely placed tubes in FIFO basis

2. Get the unplaced object frame in each tube in FIFO basis.

3. Place maximum number of object frames with no collision in synopsis frame.

4. Give more priority to the object frames from the tubes which are being generated first.

5. Remove the object frame from the tube if it is being placed in the synopsis video.

6. Remove the tubes from the cluster if all the object frames are being placed in synopsis video and add new tubes to keep the cluster size constant.

}

## 3.4 Image Enhancement

*3.4.1 General Image Enhancement*

Fig.3.6 indicates the image enhancement functionalities. The user could pause the video and draw a region of interest to be cropped out. For cropping, from the top left a submatrix of the drawn region is created and displayed. A reset button is included to undo incorrect cropping by displaying prior image from memory.

Then image enhancement options could be adjusted based on the preview and user preferences.

- Noise reduction – Image is initially separated to YCbCr color space channels. Depending on the kernel size σ given by the user through the slider, Gaussian filtering is carried out on the Y plane for general noise reduction and Cr and Cb planes for color noise reduction. The combined channel RGB image is then displayed as the output.

- Hue, Saturation, Value (HSV) – The image is converted from RGB color space to HSV color space. The user can adjust each channel independently by separate sliders. The combined output is displayed.

- Luminance – RGB image is converted to YCbCr (Luminance, Chroma blue, Chroma red) and then split. The Y channel is adjusted according to user input slider value and merged prior to display.

- Shadow/Highlight – RGB channels of the image are split and a customized Gamma correction criterion is applied to each channel based on user given slider values.

- Sharpening – The user provides the Gaussian filter kernel size and a weight value through sliders. The former is used to Gaussian blur each RGB plane of the image. The latter is multiplied with the blurred image and then this is subtracted from the original to obtain the sharpened output image.

- Contrast – User adjusted slider value is in the range 0 to 2. Values greater than 1 increase contrast. This represents dark pixels becoming blacker and light pixels whiter. Auto-contrast adjustment is given by dividing 255 by the difference between the maximum and minimum intensity values in the image.

- Exposure – Exposure is the combination of contrast and brightness. The value from the contrast adjustment slider is multiplied with the image and the exposure slider value is added back in a linear transformation manner to give the output image.

- Temperature – Warmer implies higher red intensity and lower blue, while cooler depicts vice versa. The temperature level is adjustable by a slider. The image is split to RGB channels and cubic spline interpolation is applied to R and B planes to extend the intensities accordingly. After merging back, the image is split once again to HSV color space and the S layer is interpolated as before. The combined image is finally displayed as the output.

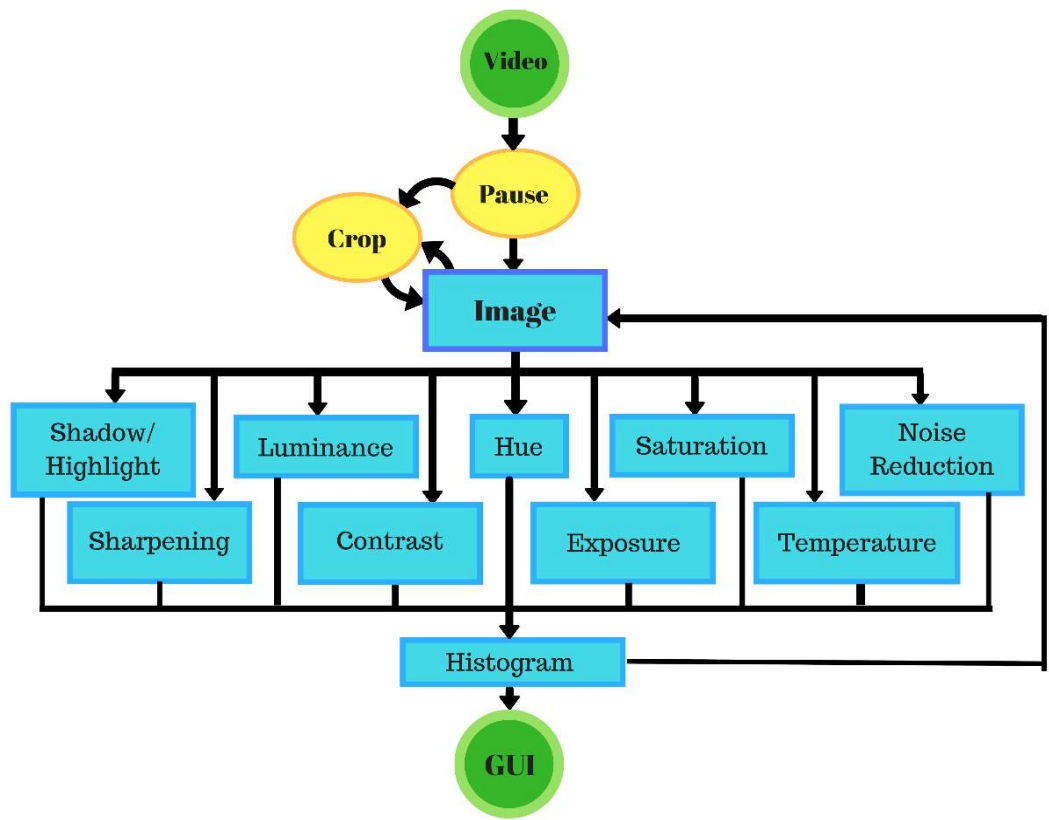

Fig.3.6 Image enhancement procedure

Further the RGB histograms facilitate decision making by quantitatively representing pixel intensities of each plane in the range of 0 to 255. Left region of the histogram (0) represents black and shadows whereas towards the right (255) indicates whites and highlights. The mid region denotes mid-tones. Finally, the user can save the enhanced image.

Histogram equalization is another method applied for enhancement of detected objects.

*3.4.2 Face Enhancement*

Since most CCTV output have low resolution, many suspect faces would be blurred or unidentifiable with the naked eye or through general image enhancement techniques. Hence a super-resolution technique that analyses several consecutive frames to develop a clear image of faces is used.

The three input parameters are

- Beta – asymptotic value of steepest descent method

- Lambda – weight parameter to balance data term and smoothness term

- Alpha – parameter of spatial distribution in Bilateral Total Variation (BTV)

First the initial image is magnified, and linear interpolation is carried out to resize. Then the image is converted to a vector to apply steepest descent method for L1 norm minimization with BTV. L2 norm is also a possibility but has lower performance. At the end each vector is converted back to matrix format prior to display. The number of iteration is given by the user. Fig.3.7 Indicates face enhancement process used.

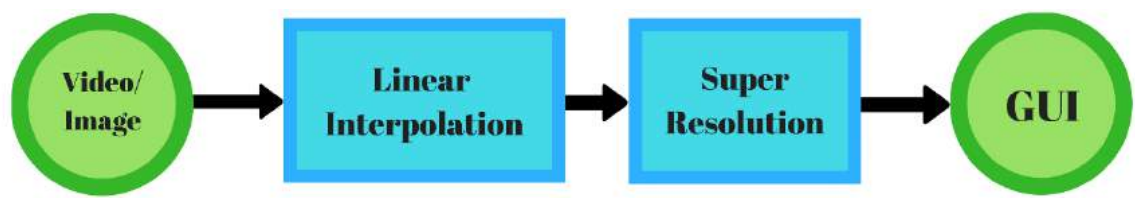

Fig.3.7 Face enhancement procedure

Another method developed, yet not integrated includes the steps of cropping and zooming into the region of interest within a frame, deinterlacing to remove small horizontal lines, frame aggregation and averaging to increase resolution, denoising, light and color adjustments, contrast adjustment, sharpening and deblurring.

*3.4.3 Textual Enhancement*

Though there are several approaches to remove the motion blur and gaussian noise, none of them give satisfactory results as deconvolving with wiener filter algorithm. Here the user estimates the point spread function by estimating the linear motion of camera by "len" pixels with angle of "theta" degrees in a counter clockwise direction by moving sliders. To remove gaussian noise the user estimates the noise variance by moving slider. Noise to signal power ratio and point spread function is given as input to deconvolve the image using wiener filter algorithm.

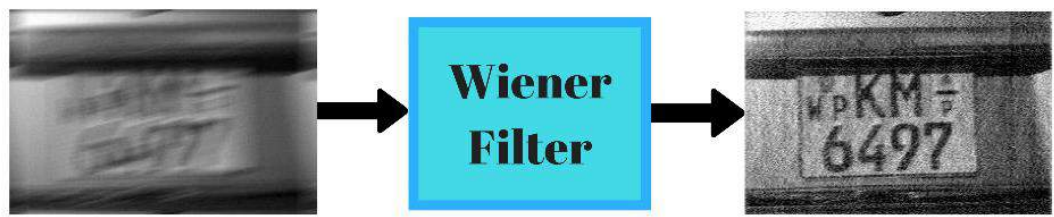

Fig.3.8 Textual enhancement procedure

**3.5 Tampering Detection**

It is very useful to analyze videos and images for their clarity since there has been growing concern for checking the reality of the images and videos which are exchanged over social media and web-based media. Also checking whether footages are real or edited is important in crime investigation for law enforcement agencies. We have categorized tampering detection into three major types which are,

- Inter-frame tampering detection,
- Intra-frame tampering detection and
- Camera tampering detection

where the first two fall under forensic and the camera tampering detection is related to surveillance.

*3.5.1 Inter-Frame Tampering Detection*

A given video is checked for frame tampering which are namely frame insertion, deletion and duplication. We used a similar method to [27] with some of work being modified. We used dense optical flows using Farneback algorithm. Once the total optical flow values have been calculated for each frame the optical flow variation sequence is built. Based on the assumption that any inter frame forgery will introduce discontinuities in the optical flow variation sequence the algorithm finds anomalies by modeling the optical flow variation sequence as a gaussian model. We also identified a research outcome by adding optical flow values, optical flow variation values and velocity field values as features to identify inter frame forgery as an anomaly detection problem. Following figure describes the approach for forgery detection

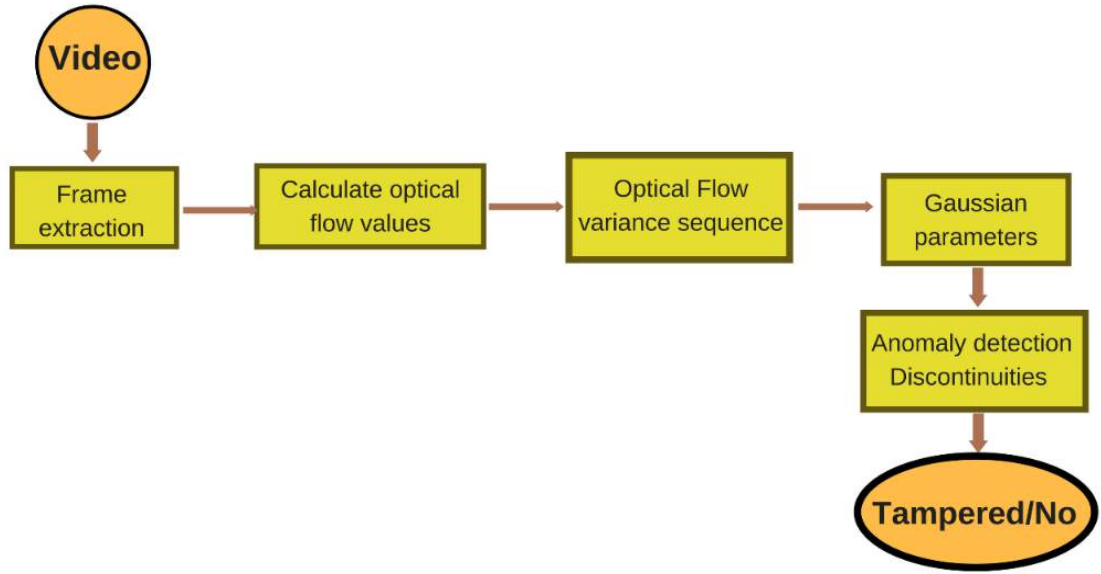

Fig.3.9 Inter frame forgery detection procedure

*3.5.2 Intra-Frame Tampering Detection*

The first intra-frame technique we discuss is copy-move forgery detection. This forgery represents copying image portions and pasting in the same image covering another area. We implemented the method in [32] which uses SIFT (Scale Invariant Features Transform) features. Given an image SIFT features are calculated to each points and the points are matched based on the SIFT values. Once the algorithm derives at matched points those are clustered in order to find copied and moved regions. Finally the algorithm estimates the possible geometrical transformation of the copied and moved areas. Following figure describes the algorithm explained.

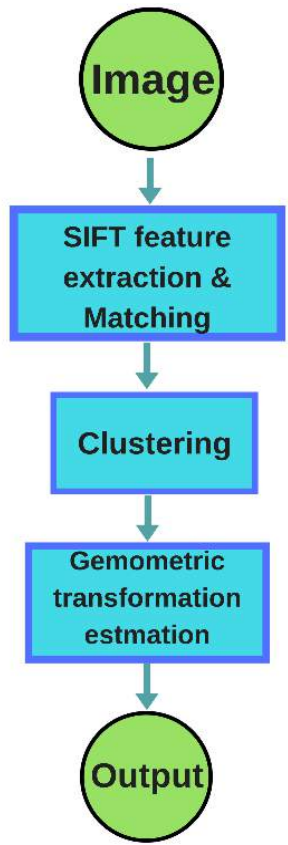

Fig.3.10 Copy move forgery detection procedure

Second technique is for splicing detection which focuses on the double quantization effect of JPEG compression. Once an image is given the algorithm first calculates the DCT coefficients and quantization matrices. DCT coefficients of all the blocks are then gathered together to build a histogram. Using these histograms, the system computes probabilities of each block being tampered which are then accumulated to derive at block posterior probability map(BPPM) which is then thresholded to identify possible tampered regions.

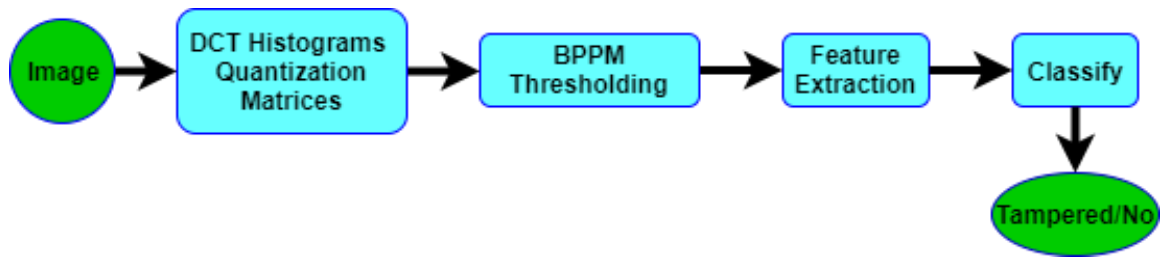

Fig.3.11 Splicing detection procedure

*3.5.3 Camera Tampering Detection*

Surveillance footages are processed in real time to detect camera shaking, redirection and camera block. We implemented a simple algorithm which extract the foreground of frames to derive at the moving objects. The total contour area of the foreground is calculated and if the portion of total foreground contour area compared to the total area of the frame exceeds a threshold the system alarms that camera tampering is being taken place. This threshold must be tuned according to the setting of the view which is captured by the camera. Figure below shows this process.

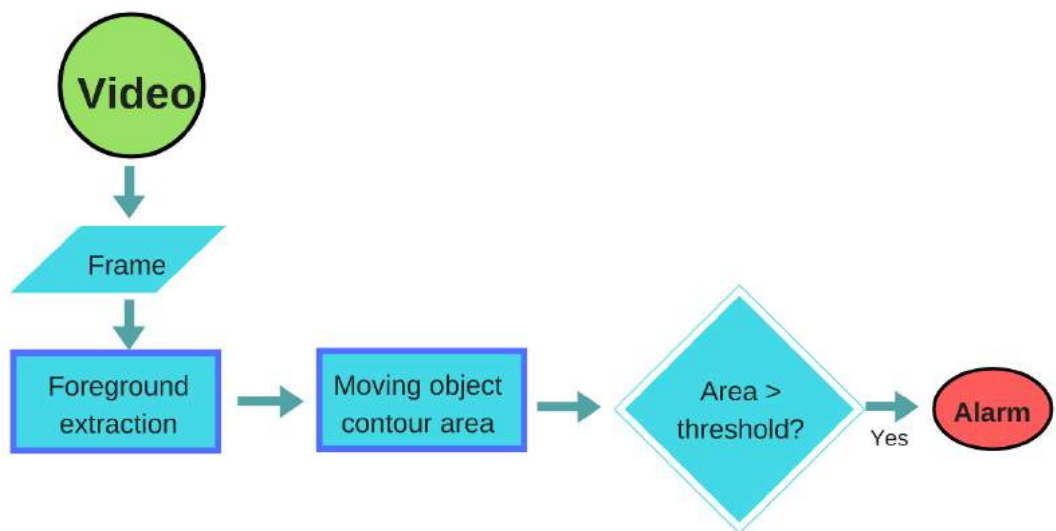

Fig.3.12 Camera tampering detection

**3.6 Object Detection**

Object detection involves detecting the bounding boxes of the objects in a given image and recognizing those objects. We used YOLO (You Only Look Once) state of the art object detector in our implementation [8]. The current evolution of YOLO is YOLO V2 [11] which has some improvements over the previous implementation with regards to mean average precision without reducing the detection rate. This is a unified architecture where bounding box detection and object recognition are being done parallel together which makes YOLO the fastest existing object detector.

The system divides the input image into an S x S grid and each grid cell predicts B number of bounding boxes of which the center falls onto that cell. It also predicts the confident scores of the box containing and object. In addition to that each grid cell also predicts the conditional probability of the object being one of the predefined classes given an object is there. At test time these confident scores and probability scores are multiplied which gives class specific confidence scores for each box. Extracted from [9] the following explains the basic functions of YOLO.

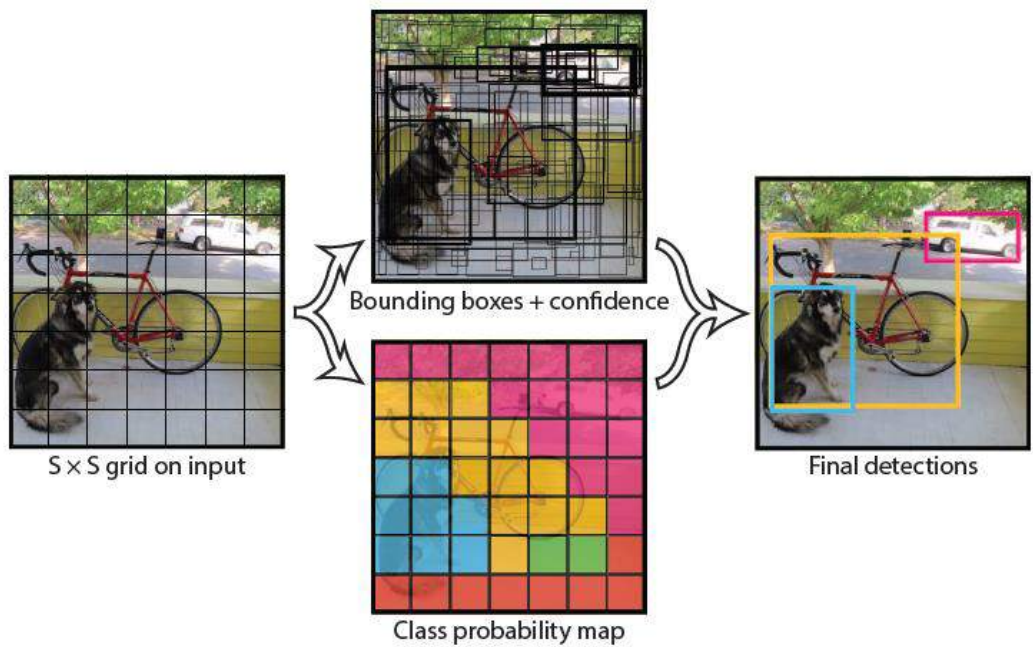

Fig.3.13 Object detection process [9]

The object detector was used as a supplementary block for producing multiple object tracking and selective object enhancement.

**3.7 Multiple Object Tracking**

Simple Online and Realtime Tracker (SORT) [14] in C++ was used to track 2D multiple objects. This is a tracking-by-detection technique that uses the bounding box co-ordinates and prediction confidence values of detected objects. Hence the tracking accuracy is can be improved as much as 20% based on detection accuracy.

Online refers to the usage of only current and past frames for tracking. This is required for surveillance due to speed and real-time requirements.

Multiple objects are detected using YoloV2 and the bounding box details with prediction confidence is fed into the tracker. Only objects with probability exceeding 0.5 are considered. Kalman Filter is used for motion estimation and each target is assigned a vector of the bounding box area and the position and velocity of the center point x, center point y and bounding box scale of each object.

Hungarian algorithm is used to solve the data association problem of assigning detection to an existing target. Here an assignment cost matrix as the IOU distance between each detection and all predicted bounding boxes from existing targets is solved optimally. Further if detection/target overlap is less than another parameterized minimum IOU, it is rejected. This also handles short term occlusion. A track is terminated if it is not detected for a certain number of frames preventing localization errors.

A new track Id is assigned if the overlapping region is less than a predefined minimum Intersection over Union (IOU). The created track is initialized with zero velocity and a large covariance due to uncertainty. Further there is a probationary period to avoid tracking false positives. Figure 3.14 demonstrates the tracking procedure.

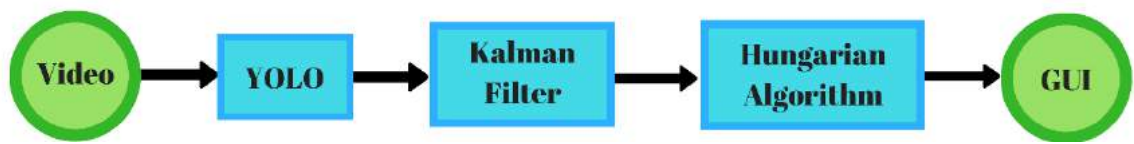

Fig.3.14 Multiple object tracking flow chart

**3.8 Anomaly Detection**

The procedure for anomaly detection [20] is as follows.

Training:

1. Sparse optical flow parameters are calculated for extracted features.

2. Histogram of optical flow is calculated with 9 bins and feature matrix is created.

3. After processing a feature matrix vector of length 5, maximum of feature matrix is saved to busyness matrix (thus for each block (8x8 pixel) and each bin (360 degrees in 9 sections of each 20 degrees)).

Testing:

1. During the testing the procedures from 1-3 are followed.

2. If the calculated busyness matrix has deviation form saved matrix form training session they are indicated as anomalies and alarm in the GUI is raised.

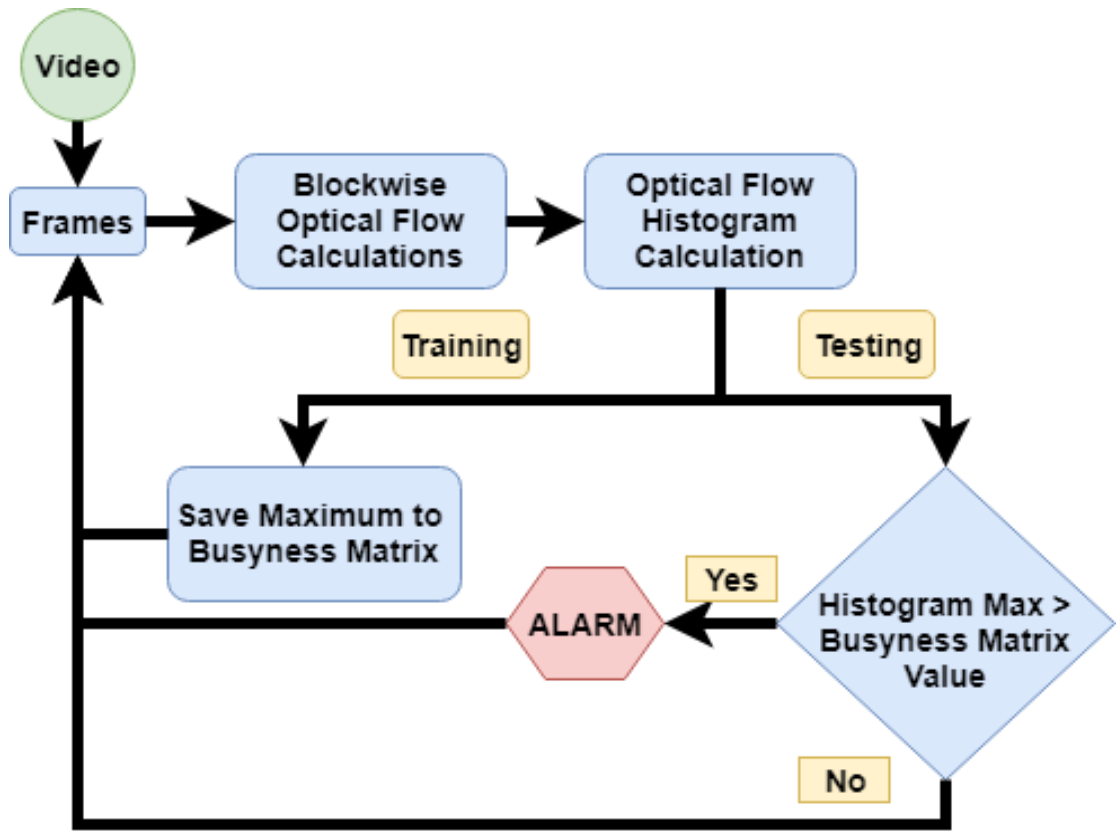

Fig.3.15 Anomaly detection flow chart

## 3.8.1 Trespass Detection

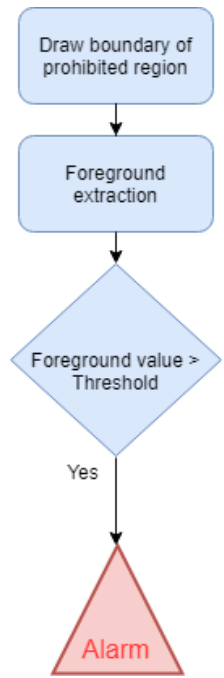

Fig.3.16 Trespass detection flow chart

Trespass detection implementation is based on foreground extraction.

**3.9 Activity Recognition**

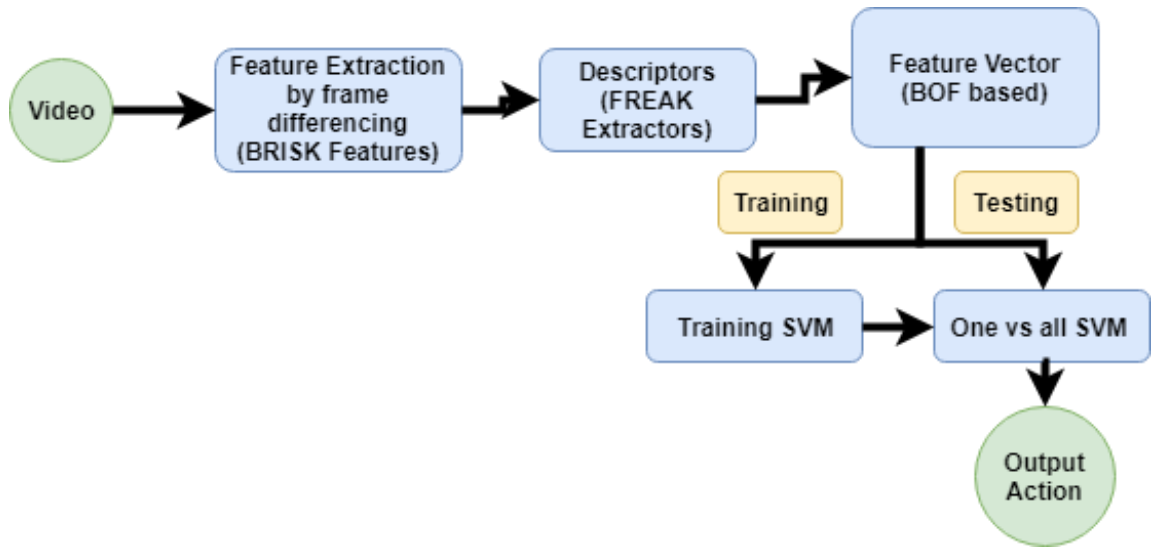

Fig.3.17 Activity recognition flow chart

We recognize different human activities both interactive and individual by training the features of labelled activities with different datasets using SVM. We use BRISK features and FREAK [15] extractors and create a BOF [16] representation for the activity video and extracted features are classified based on one vs rest SVM. Thereby, we can recognize activities like fighting, robbing, meeting, hugging, unusual running, following. Here as activity recognition was developed for surveillance need this must be fast and real-time and we concentrate mostly on surveillance related erroneous human activities only.

Localization of activity recognition is done by thread based multiscale video processing [45]. Frame is divided into multiscale overlapping and non-overlapping regions. Thus, localization is not an additional computational complexity due to threading. As in below figure localization of activity recognition is done by overlapping and non-overlapping multi-scale frame processing based on maximum feature point region of interests.

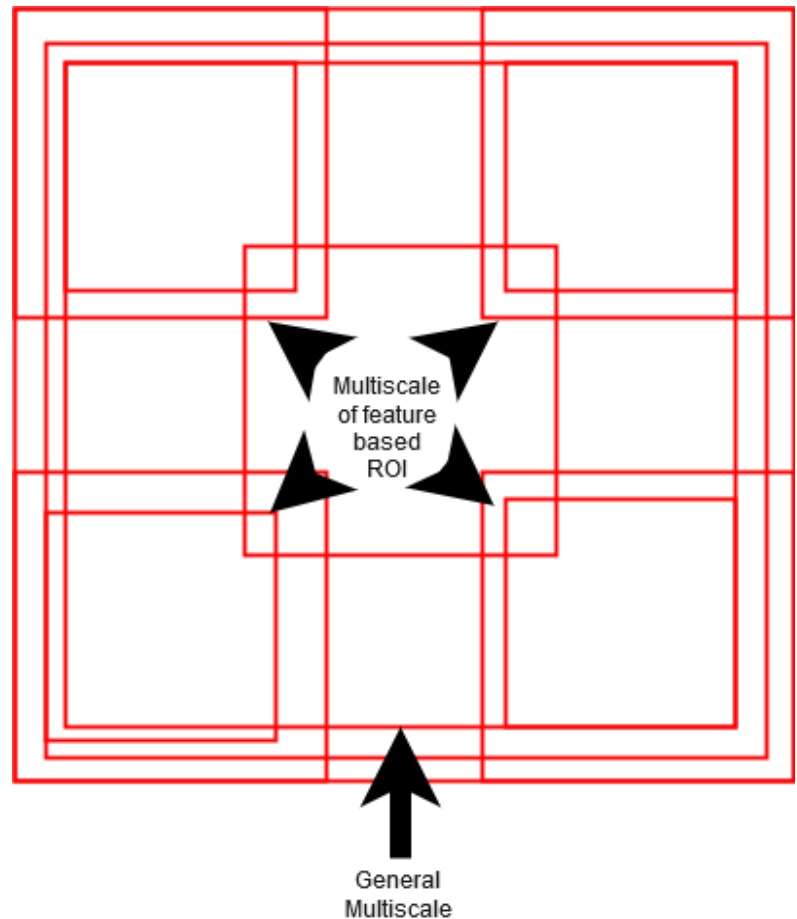

Fig.3.18 Localization of activities

Although we were able to develop activity recognition as a standalone application we were unable to integrate with surveillance application due to computational complexity of activity recognition won't allow the application to run in real-time. In, future we would be able to integrate activity recognition with the existing application.

## 3.10 GPU

GPU is utilized to speed up training of neural networks and classification for YOLO based object detection. Anomaly detection mainly relies on CUDA OPENCV library for speeding up its processes. We are currently using NVIDIA GTX 1050Ti 4GB, for object detection and anomaly detection. Resource usage and possibility of performance enhancement and optimizations are discussed under Results chapter.

**3.11 Reference Tools**

The existing security tools are expensive, equipment specific, region based and have complex maneuvering with a large learning curve. The following are such tools analyzed prior to development.

- BriefCam
- Milestone
- AGENT Vi

**3.12 Sri Lankan Context**

Considering the Sri Lankan context is a critical factor of this software since our ultimate goal is to use this software in Sri Lanka. We have identified several applications of this software in the Sri Lankan context. Sri Lanka already has some CCTV cameras mounted to monitor traffic and more CCTV cameras will be mounted to monitor traffic. With the help of this software it will be more helpful to detect traffic violations and identify the vehicles where necessary. In addition, we can use the software to detect abnormal behaviors of people in order to detect violations. For example, we can take unauthorized railway crossings. Shop theft has also been one of the major crimes taking place in Sri Lanka. This software will be able to identify such incidences by identifying anomaly behaviors and alarm without the intervention of people. The Sri Lankan Police is facing difficulties with finding evidences from CCTV video feed which recorded the crime scenes due to the quality of those videos. We can use this software as an evidence extracting tool for such occasions. With the fast-growing usage of social media and other web-based media we can see tons of information is being shared. It is useful to check whether the visual content which is being shared is real or edited. Also tampering detection will be useful for law enforcement agencies to check the clarity of the footages and photos. Likewise, in the Sri Lankan context there is a need and a demand for this kind of software which will be very helpful in both surveillance and forensic aspects.

**3.13 Add-on in Milestone**

Milestone is a vast video managing software platform that issues their customized video analytic software as well as APK to develop add-ons for their software. It is a way to realize and market software into the real world. Developing add-ons for their platform which can be downloaded with their software had many restrictions. A minimum requirement of Windows 10 Pro constrained our software development and would prohibit applicability among small scale users especially in Sri Lankan context. A registered company was also required to create add-ons and obtain Milestone certification. Hence for the scope of the project integrating our software with Milestone was annulled.

**3.14 Development**

The development tools used include the following.

- OS – Microsoft Windows

- Programming languages – C++, Matlab

- Libraries – OpenCV 3.4.0, Cuda 8.0, CuDNN 5.1

- Platforms – x64 Visual Studio 2015, Matlab

- Software frameworks – .NET framework v4.5.2

- GPU – NVIDIA 940MX, NVIDIA GTX 1050 Ti

- GPU Toolkits – Cuda 8.0, CuDNN 5.1

- CPU – Intel® core™ i7-4770 @ 3.4GHz, Intel® core™ i7-7500U @ 2.70GHz, Intel® core™ i7-7700K @ 4.20GHz

- RAM – 4GB, 8GB, 16GB

**3.15 Summary**

This chapter demonstrated the methodologies used under each category of the project namely multiple object detection, multiple object tracking, anomaly detection, activity recognition, image enhancement, tampering detection and video synopsis along with architecture and development tool details. The market need for our software is also considerable in the Sri Lankan context.

# Chapter 4

## RESULTS

This chapter presents the results of both forensic and surveillance blocks. Video Synopsis research paper outcomes are provided with in-depth analysis of results with comparison of parallel research contributions. Popular datasets have been used for this purpose. Further Activity recognition, image enhancement and camera tampering outcomes have been depicted. Applications and Sri Lankan contextualization of results are analyzed in this section.

**4.1 Video Synopsis**

To evaluate the performance of the proposed video synopsis algorithm, video datasets in [44] and [2] have been used for evaluation. As the accuracy of tube generation depends on how well the proposed method detects and tracks multiple objects, tracking and vehicle counting accuracy within the region of interest of the annotation of the GRAM dataset [44] have been evaluated below.

Detailed information about the videos used for tracking and vehicle counting is as follows.

Table.4.1 Detailed information about GRAM dataset

| Video Name | M-30-HD | M-30 |
|---|---|---|
| Size of the image (96 dpi) | 1200*720 | 800*480 |
| Total Number of Vehicle | 241 | 270 |
| Frames per second | 30 | 30 |
| Total Number of Frames | 9310 | 7520 |
| Weather Condition | Cloudy | Sunny |

We have used the MATLAB code given by GRAM dataset [44], to evaluate the accuracy of the tracking algorithm. The above code calculates the average precision to determine the accuracy of tracking algorithm and it plots the "precision" vs "recall" curve. The definition of "precision" and "recall" used here are as follows.

Let,

TP[n] – True positive in n<sup>th</sup> frame

FP[n] – False positive calculated using false detection and multiple detection in n<sup>th</sup> frame

NP[n] – Total number of annotated detection in n<sup>th</sup> frame

N – total number of frames in the video

i - i<sup>th</sup> frame

$$\text{precision} = \frac{\sum_{n=1}^{n=i} TP[n]}{\sum_{n=1}^{n=i} TP[n] + FP[n]} \quad (4.1)$$

$$\text{recall} = \frac{\sum_{n=1}^{n=i} TP[n]}{\sum_{n=1}^{n=N} NP[n]} \quad (4.2)$$

$$\text{average precision} = precision * recall \quad (4.3)$$

Table 4.2 shows comparison of average precision values calculated using different multiple object detection and tracking approaches with the proposed approach.

Table.4.2 Average precision of existing and proposed methods

| Detection Method | Tracking Method | M-30-HD | M-30 |
|---|---|---|---|
| **Multi Time Spatial Image Based Vehicle Detection [2]** | KF | 0.478 | 0.291 |
| | PF | 0.681 | 0.664 |
| | MIKF | 0.806 | 0.741 |
| | MIPF | 0.769 | 0.701 |
| **HOG [41]** | Extended Kalman Filter | 0.524 | 0.3009 |
| **Proposed Method** | **Proposed Method** | 0.871 | 0.799 |

As the higher average precision value relates to the higher accuracy, from the above table the proposed method detects and track accurately. Fig.4.1 shows the precision vs recall curve for both videos.

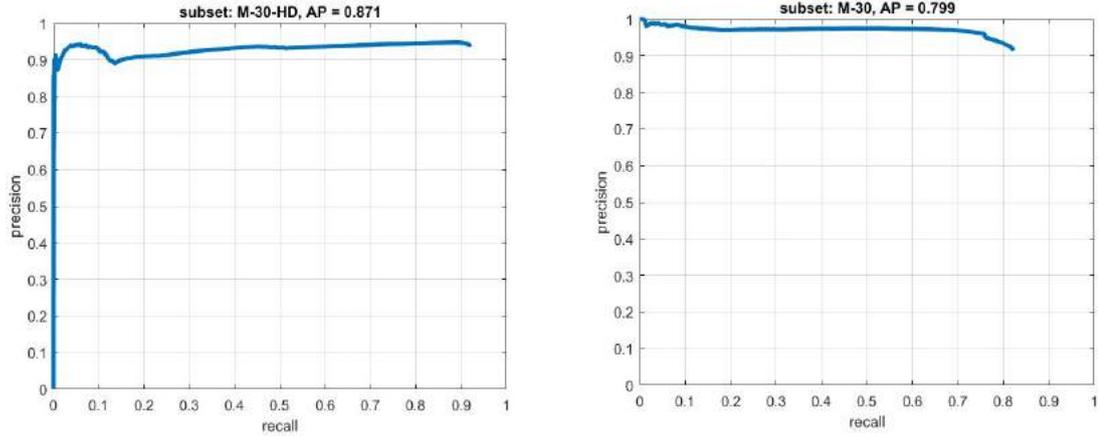

Fig.4.1 Precision-recall curve for tracking M-30-HD (left) and M-30 (right) videos

The precision recall curve for M-30-HD video depicted in the left has zero precision in the beginning. This because of false detection during the phase where the Gaussian Mixture Model starts to train to last 100 frames. After 100 frames we can see sudden increase of precision value to 90%. In both cases we have trained Gaussian Mixture Model with 1st 100 frames and we re-ran the video to get annotation. For M-30 video the glitch was not produced. From the above graph we can say that at any instance more than 90% of the total detection is true positives.

The above graphs indicate the precision of the algorithm is more than 90% and it can detect more than 80% of annotation in M-30 video and more than 90% in M-30-HD video accurately.

The specialty of proposed approach is that, it can detect and track larger region of interest (ROI) than the annotation of the above video. Fig.4.2 depicts that proposed approach detect and tracks highly accurate within the ROI of annotation while it also can detect and track outside the ROI at acceptable accuracy.

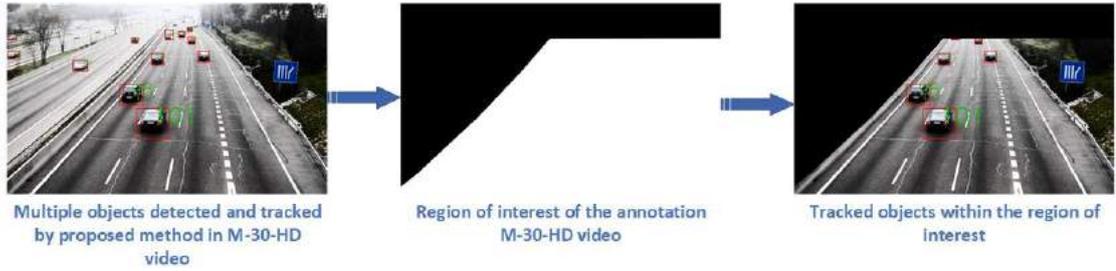

Fig.4.2 Multiple detection and tracking of 1000th frame (left), ROI of annotation (middle), multiple detection and tracking within the ROI(right)

The video synopsis algorithm has been tested on the 4 videos used in [2]. The detailed information of the dataset is as follows.

Table.4.3 Detailed information about synopsis dataset

| Video | Duration | # of frames | FPS | # of objects |
|---|---|---|---|---|
| **Cross Road (V1)** | 01:04:59 | 70195 | 18 | 1677 |
| **Street (V2)** | 01:13:33 | 79449 | 18 | 871 |
| **Hall (V3)** | 01:01:49 | 66771 | 18 | 276 |
| **Sidewalk (V4)** | 00:58:18 | 104864 | 30 | 334 |

We have experimented on different cluster size (CS) of the tubes and produced the following results.

Table.4.4 FPS and FR for different cluster size for dataset videos

| CS | Cross Road | | Street | | Hall | | Side Walk | |
|---|---|---|---|---|---|---|---|---|
| | FR | FPS | FR | FPS | FR | FPS | FR | FPS |
| **10** | 0.211 | 56.9 | 0.213 | 43.2 | 0.121 | 85.9 | 0.120 | 89.3 |
| **20** | 0.167 | 58 | 0.141 | 43 | 0.088 | 85 | 0.095 | 87.3 |
| **40** | 0.142 | 56.6 | 0.112 | 42.9 | 0.073 | 83.7 | 0.084 | 84.3 |
| **200** | 0.121 | 55.1 | 0.095 | 40.5 | 0.07 | 67.4 | 0.078 | 72 |
| **1000** | 0.119 | 46.2 | 0.095 | 35.1 | 0.07 | 57.5 | 0.077 | 56.2 |

Here Frame reduction rate(FR) is calculated by dividing total number of frames in synopsis video to the total number of frames in original video. Frame per second (FPS) is calculated by dividing total number of frames in original video to the total time taken to create synopsis video.

Through the experiment it can be concluded that time taken produce synopsis video is proportional to synopsis video size and average object density in original frame while it is inversely proportional to the cluster size. Although frame reduction rate is inversely proportional to the cluster size, the synopsis video becomes unpleasant to watch at large cluster size since there will be more flickering in video to avoid occlusions and all the objects will be tightly packed in the video.

After the experiment optimal cluster size for four videos were chosen based on the creation of visually pleasing synopsis video, frames per second and frame rate. The proposed method has been compared with existing methods which have used the above dataset and following results have been produced.

TOV – Total number of frames in original video

TSV – Total number of frames in synopsis video

Table.4.5 Comparison of number of synopsis frames and frame reduction rate

|  | TOV | [3] | | [2] | | Proposed Method | | |
| --- | --- | --- | --- | --- | --- | --- | --- | --- |
|  |  | TSV | FR | TSV | FR | TSV | FR | CS |
| V1 | 70195 | 12685 | 0.181 | 7876 | 0.112 | 12906 | 0.184 | 15 |
| V2 | 79449 | 18703 | 0.237 | 21371 | 0.269 | 16947 | 0.213 | 10 |
| V3 | 66771 | 14379 | 0.215 | 11271 | 0.169 | 7311 | 0.11 | 12 |
| V4 | 104864 | 18250 | 0.174 | 17340 | 0.165 | 15399 | 0.147 | 7 |

From Table 4.5 we can see that at optimal cluster size the frame reduction rate by our algorithm is lower than existing work for last 3 videos in the table. This indicates that our algorithm can efficient reduce spatial and temporal redundancies while the produced output video is visually pleasing. From Table 4.4, we can see that it is possible to reduce the frame reduction rate of Cross Road (V1) video closer to [2] at cluster size of 1000. But the produced video will be tightly packed with more flickering effects. Since visually displeasing videos irritates the viewers, we have given priority to the quality of output video when selecting the optimal cluster size.

The proposed method has been experimented in Intel® core™ i-7-4770 CPU @ 3.4GHz. The below table shows that the proposed method runs at real time in less dense video dataset.

Table.4.6 Frames per second for different videos

| Video | Original Video FPS | FPS | Cluster Size |
|---|---|---|---|
| **Cross Road** | 18 | 57 | 15 |
| **Street** | 18 | 43.23 | 10 |
| **Hall** | 18 | 85.55 | 12 |
| **Side Walk** | 30 | 89.82 | 7 |

Another use of the creation of synopsis video is that the original video can be compressed. Since CCTV cameras records 24hrs large amount of memory needs to store it, as video synopsis compresses the video with only useful information memory can be efficiently managed. The below table show the amount of compression has been done. Below table shows original and synopsis video under H.264 compression.

Table.4.7 File size under H.264 compression (BYTES)

| Video Name | Original Video | Synopsis Video |
|---|---|---|
| **Cross Road** | 49.9M | 25.7M |
| **Street** | 41.8M | 21.5M |
| **Hall** | 26.9M | 11.3M |
| **Side Walk** | 76.6M | 16.8M |

Fig.4.3 shows the synopsis video created using 4 videos. The objects in synopsis video is labelled with the time it appears in original video.

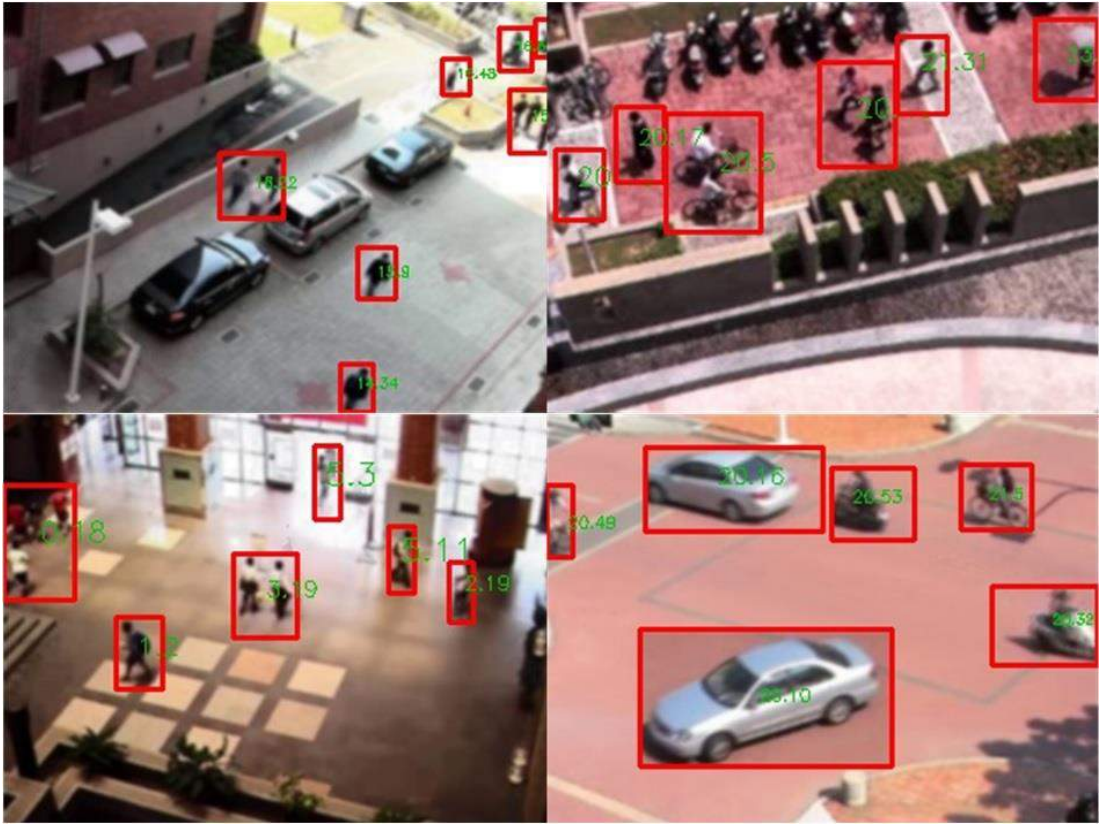

Fig.4.3 Synopsis video created using the dataset

**4.2 Surveillance Panel**

Surveillance panel performs object detection and tracking, anomaly detection, camera tampering detection and trespass detection and zooming and enhancement of region of interest. All these activities perform parallelly and were verified with video feeds from Sri Lankan context. All the monitored features in surveillance panel assumes that the camera is stationary except for camera tampering.

- Trespass detection:

Simple algorithm runs effectively. Planning to include more user-friendly boundary defining tools.

- Object detection and tracking

Implemented on two configurations,

    o On-demand object tracking of user desired objects

o   Tracking of Anomaly detected objects.

Fast and efficient as we use GPU acceleration (Yolo neural networks). But have some drawbacks, such as misses the object during an occlusion. Below figure shows the tracking after anomaly.

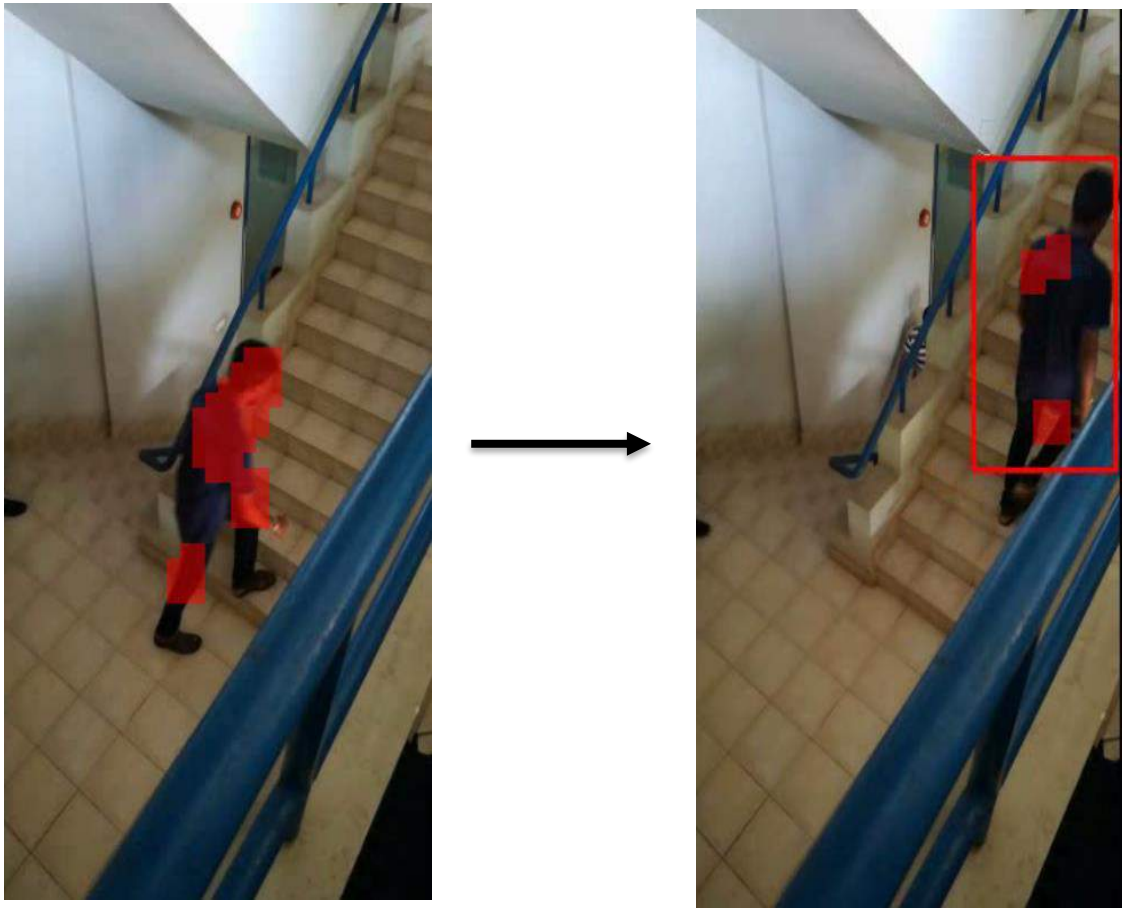

Fig.4.4 Tracking anomaly

- Anomaly detection

    Anomaly rules are defined by the train video we provide. Training must be done so that it covers all the non-anomaly parts. GPU acceleration helps to implement the process in real time surveillance. Training video length depends on the density data in each frame. The results were very much convincing for traffic as well as indoor applications. Results were checked with Sri Lankan context.

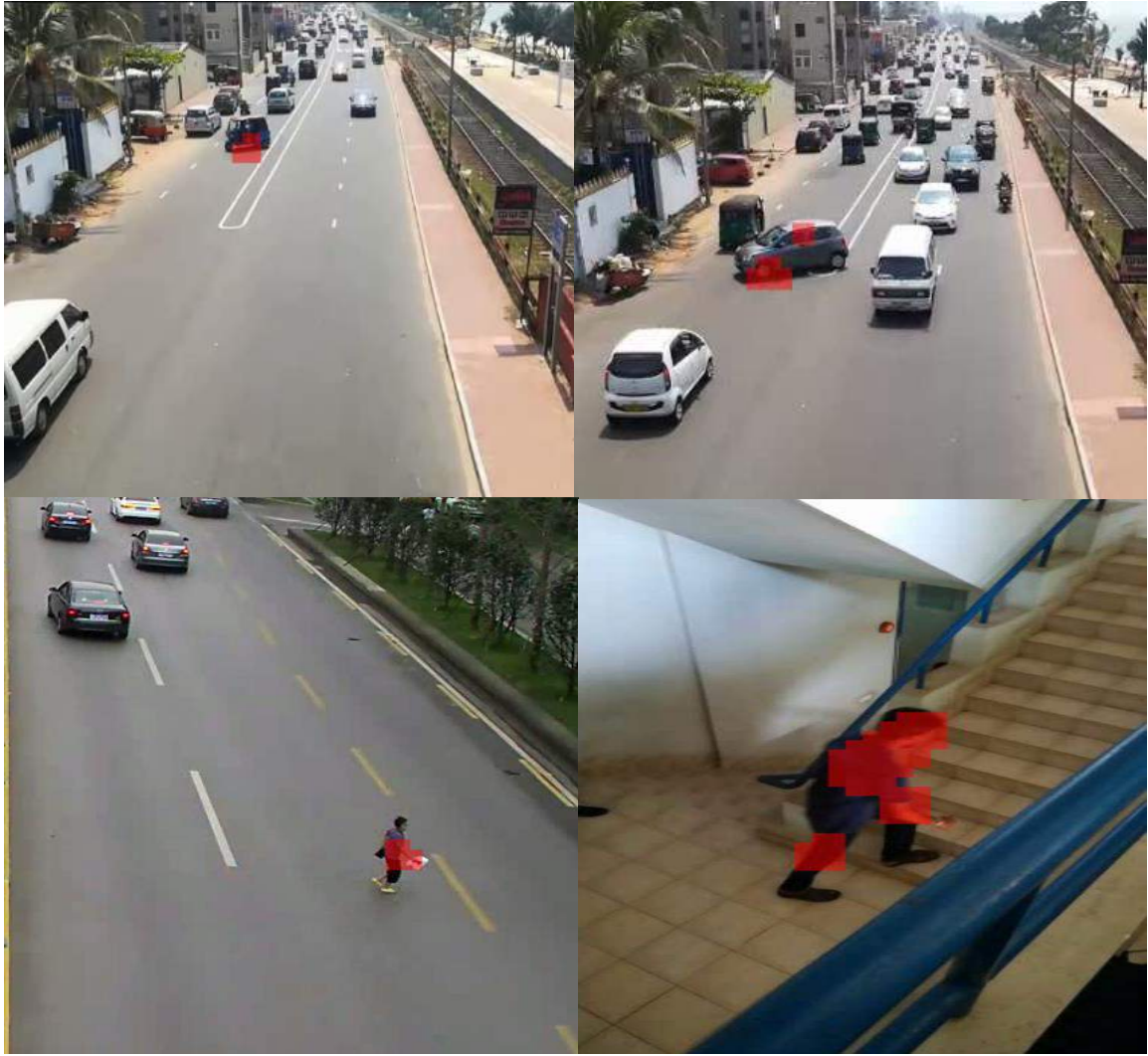

Fig.4.5 Anomalies in Sri Lankan context

## 4.3 Activity Recognition

Activity recognition part was tested with CASIA, KTH datasets. CASIA datasets provide individual as well as interactive human behaviours, while KTH dataset focus on individual activities. The module provides acceptable performance for intra dataset training and testing.

Performance analysis for activity recognition was done by, randomly perturbating videos to train and test. Following table indicates the average confusion matrix for KTH train and test (25 trials) for selected 4 activities of KTH dataset (handclapping, running, handwaving and boxing).

Table.4.8 Activity recognition confusion table

|  | Boxing | Handclapping | Handwaving | Running |
|---|---|---|---|---|
| Boxing | 0.94 | 0.05 | 0.01 | 0.00 |
| Handclapping | 0.10 | 0.87 | 0.03 | 0.00 |
| Handwaving | 0.02 | 0.00 | 0.89 | 0.00 |
| Running | 0.01 | 0.01 | 0.02 | 0.96 |

Above results were obtained without implementing localization of activity recognition. However, the frame processing speed is low so that it cannot be integrated with surveillance block. We have plans to include this, by optimizing the speed by increasing the number of threads.

**4.4 Forensic Panel**

General image enhancement is highly dependent on the user preference, creativity and analysis. Some of the results in Sri Lankan context is depicted in Figures 4.6 to 4.8.

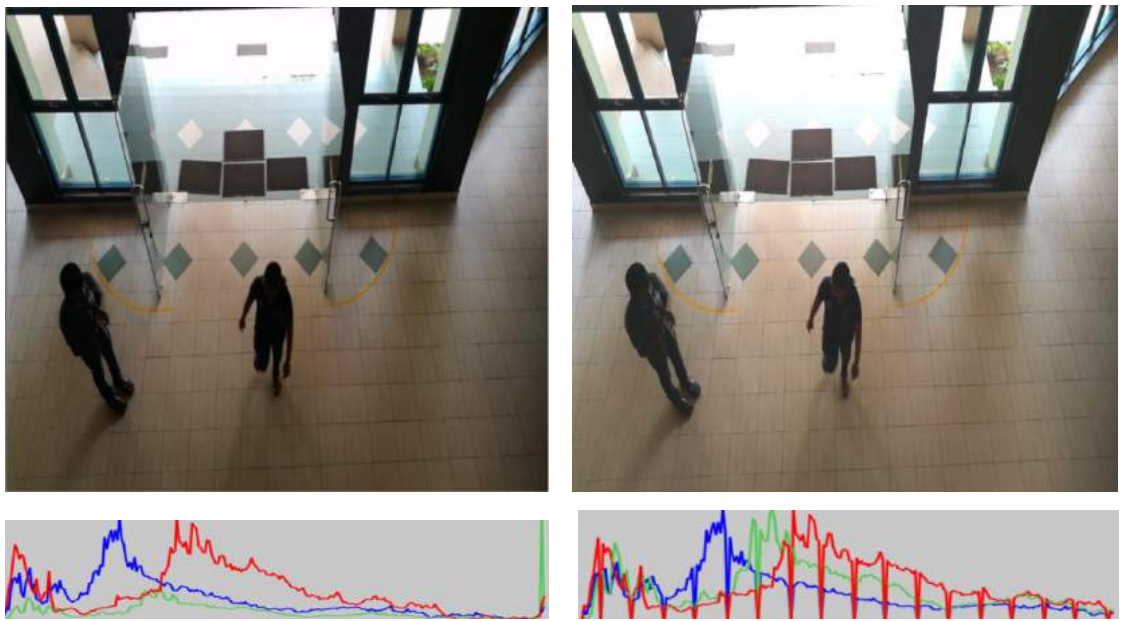

Fig.4.6 Original image and histogram (left) contrast and exposure adjusted (right)

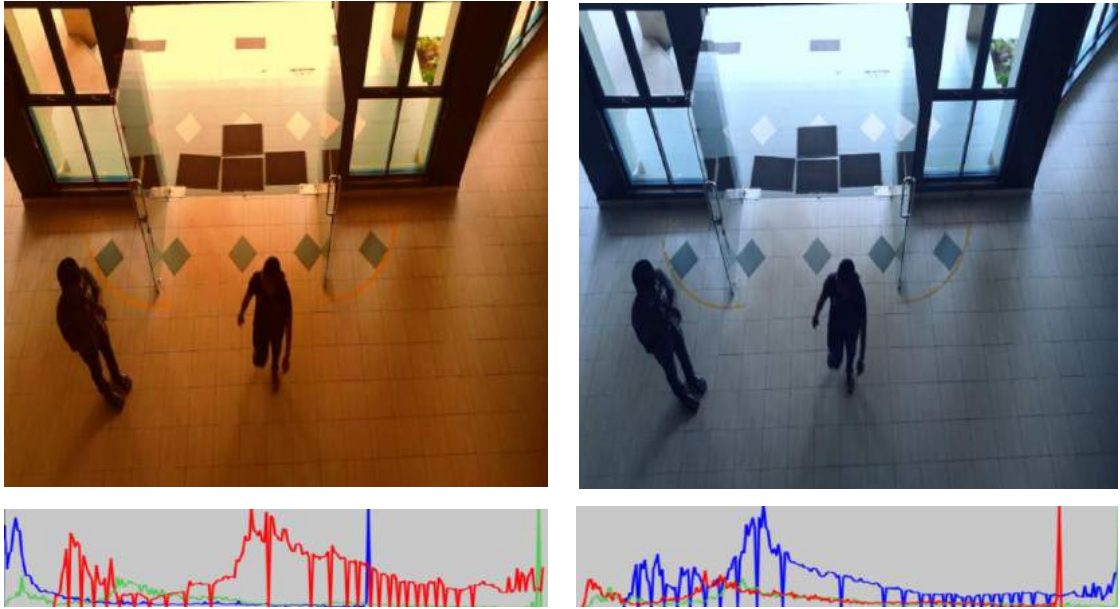

Fig.4.7 Temperature adjusted with histograms for warm (left) and cold (right)

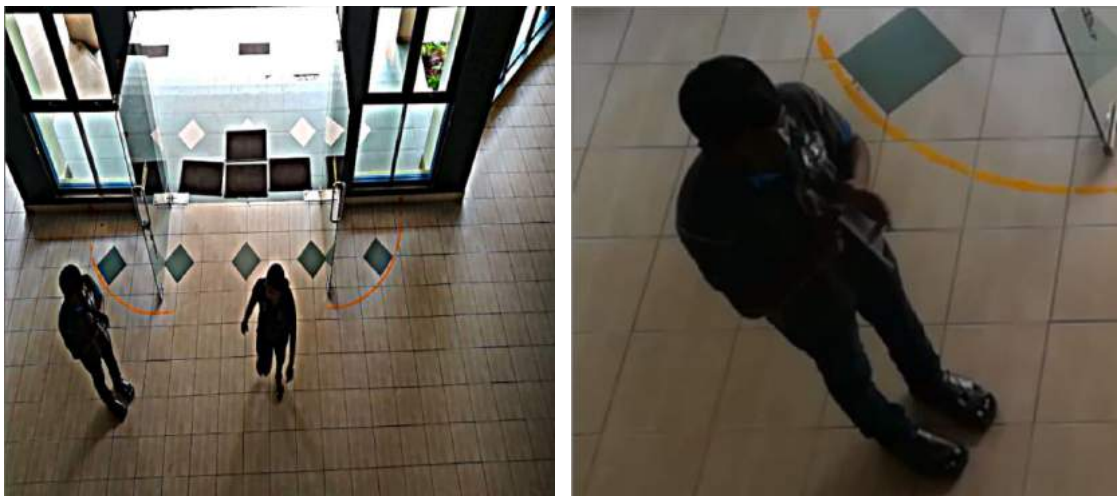

Fig.4.8 Sharpened image (left). Crop and zoom (right)

Face enhancement has been qualitatively observed for the low resolution Lenna image as shown in Fig.4.9. If Gaussian noise or Salt and pepper noise has been added to the original image, such noise can be removed successfully.

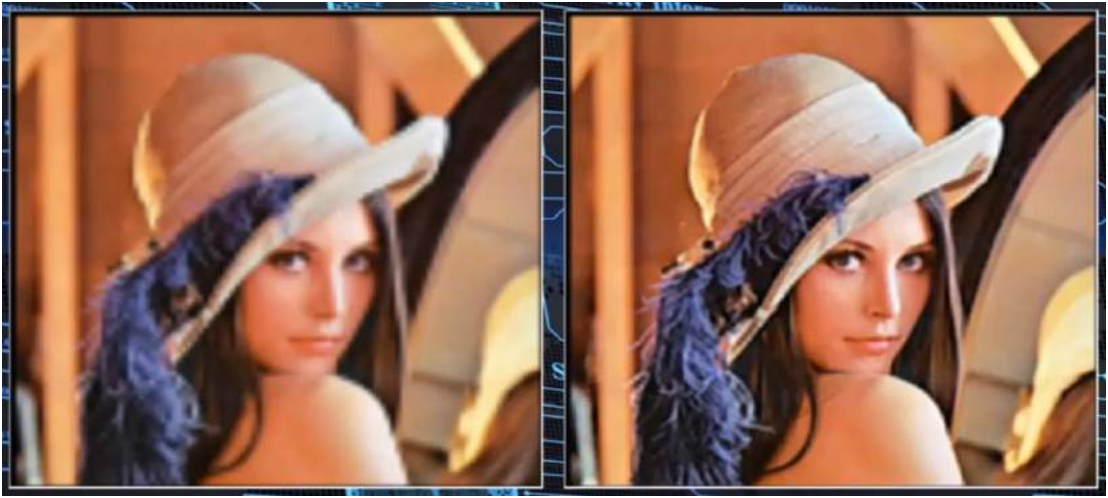

Fig.4.9 Linear interpolated image (left) super-resolution output (right)

For videos this enhancement method has produced sufficiently clear faces from consecutive frame with sufficiently large misalignment as shown in Fig.4.10.

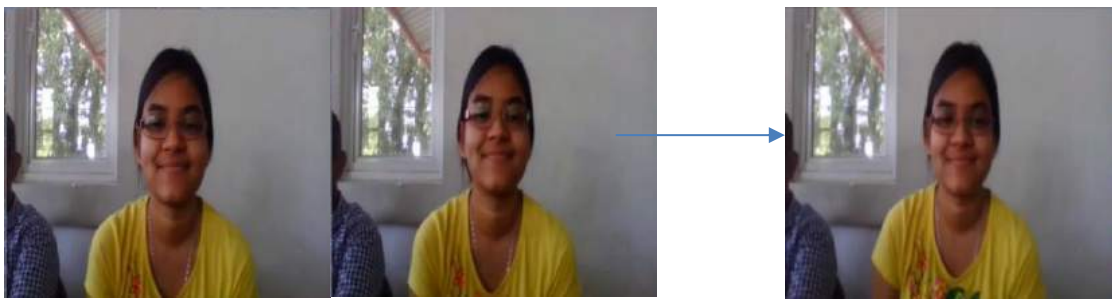

Fig.4.10 Multiple video frames with large motion (left). Enhanced face (right)

We have qualitatively evaluated the accuracy of textual enhancement by deblurring different number plates which has different level of motion blurs. Fig.4.11 shows some of the results which we have obtained.

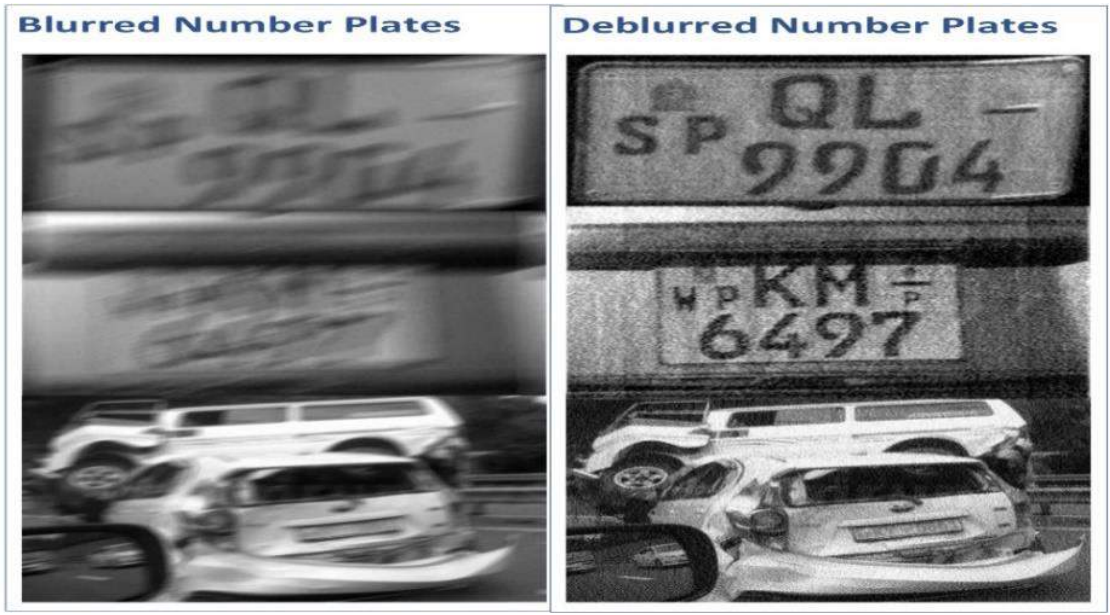

Fig.4.11 Motion blurred number plates (left) and after deblurring using wiener filter algorithm (right)

From the above results we can observe that deconvolving with estimated motion filter gives acceptable results in the 3 scenarios. This enables the user to recover the number plates, which were degraded by motion blur and Gaussian noise.

**4.5 Applications**

*4.5.1 Highways*

There are many highways in Sri Lanka covering a wide area from north to south. Although there are many CCTV cameras fixed on the highway, most of the CCTV recorded videos are shelved without being viewed. Our Video Synopsis methodology enables relevant authorities to view 24-hour video records within an hour. This will motivate the relevant authorities to have a look at the summarized videos and verify whether there has been any forensic activity. This will also ease them to track any forensic activity and collect evidence quickly.

Fig.4.12 shows a frame from the original video while Fig.4.13 shows a frame from summarized enhanced video from a highway CCTV camera.

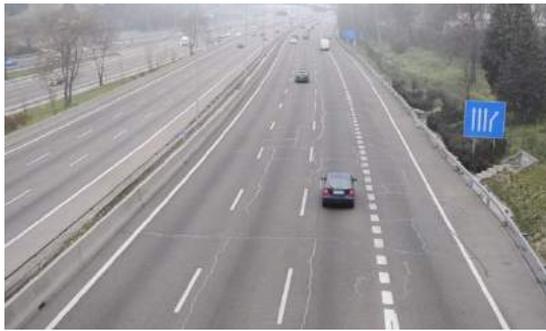 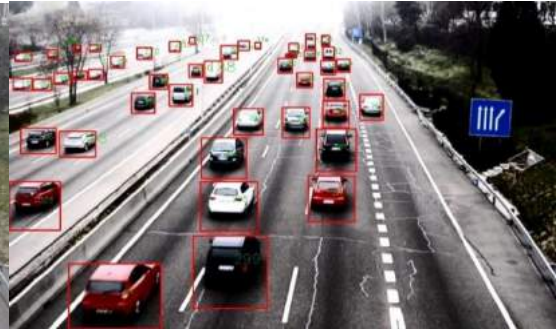

Fig.4.12 Frame from the original video      Fig.4.13 Frame from summarized video

*4.5.2 Public Places*

The work of security officers can be made easier with our video summarization solution. Using our video summary, they will be able to detect any forensic activities faster and report it to the relevant officials.

Fig.4.14 shows a frame from the original video while Fig.4.15 shows a frame from a summarized enhanced video from a CCTV camera in a public place.

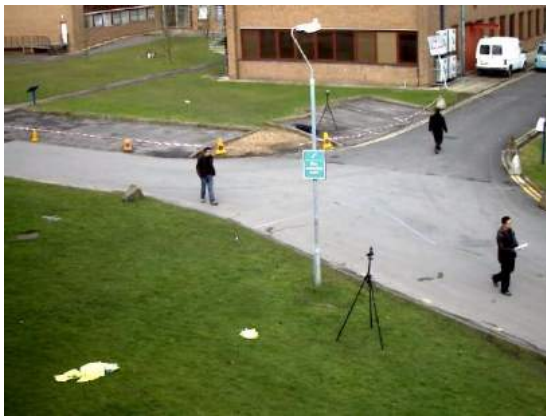 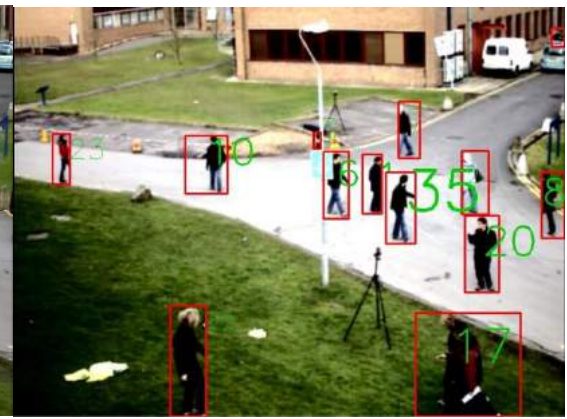

Fig.4.14 Frame from the original video      Fig.4.15 Frame from summarized video

## 4.6 Tampering Detection

The gun in fig.4.16 has been replicated. This copy move forgery has been detected and tampering alarm is raised.

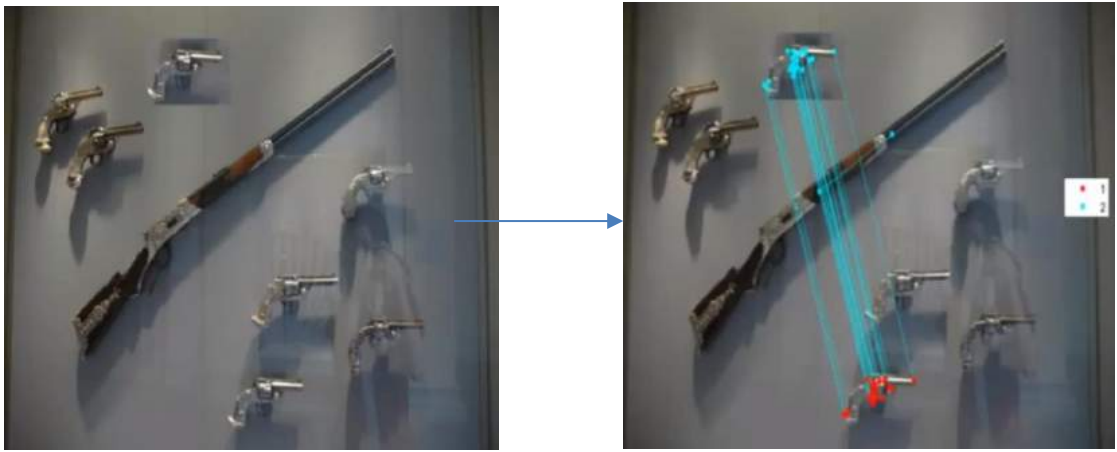

Fig.4.16 Copy move detection

The bird in the jpeg image shown in fig.4.17 has been included externally onto the background image from a different image. As a result, that region has double quantization when compressing. This is indicated through the white region as a splicing tamper.

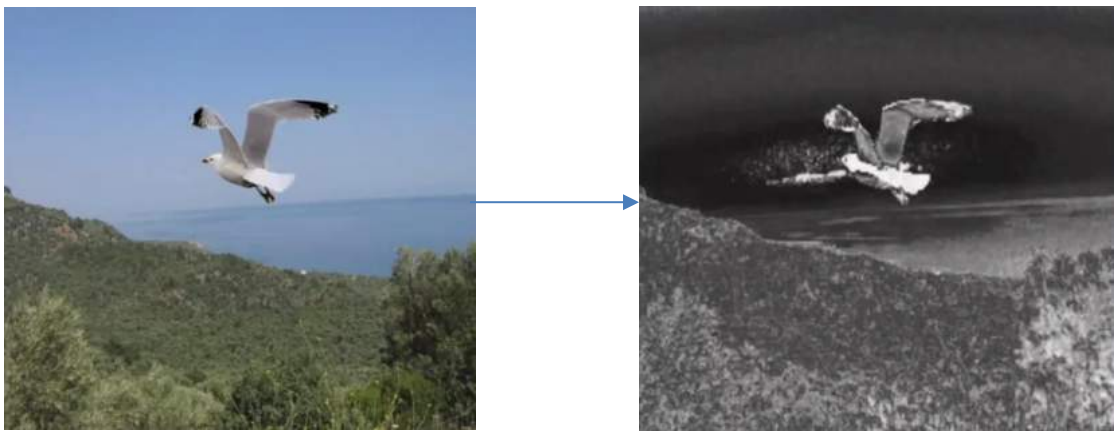

Fig.4.17 Splicing detection

Fig.4.18 shows the system detecting the same frame sequence repeatedly as inter frame forgery. The red truck passing scene is being repeated and it is detected

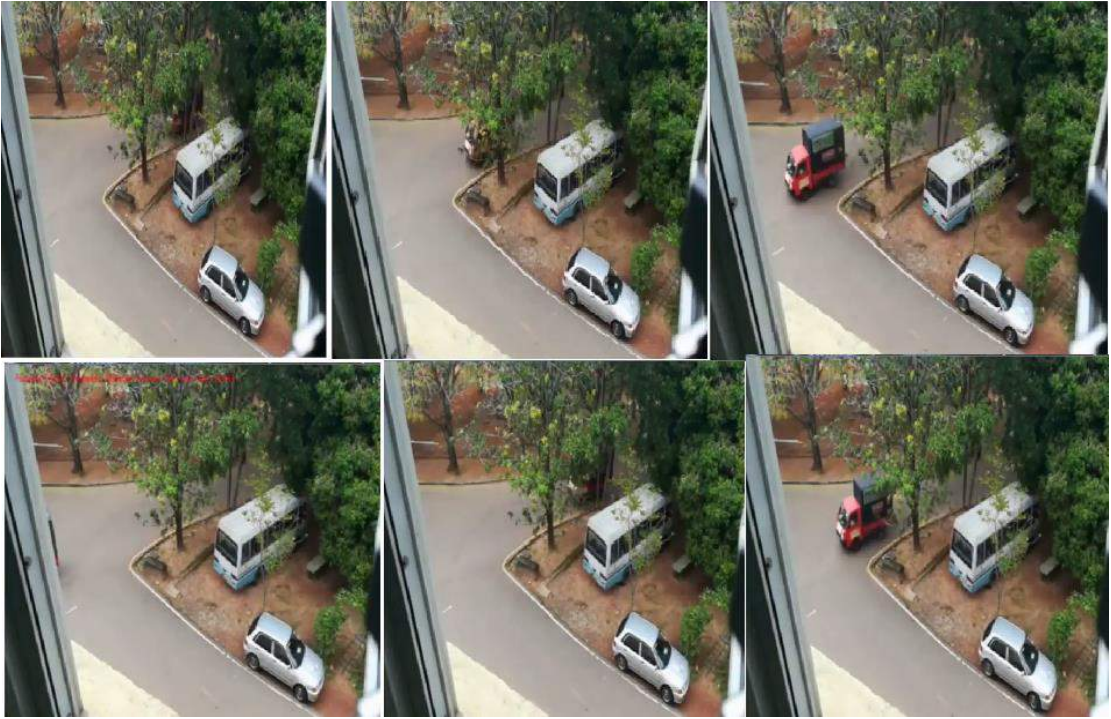

Fig.4.18 Inter frame forgery – frame repetition detection

Following figures show the system detecting camera tampering forgeries, namely are camera shake and camera block and camera redirection.

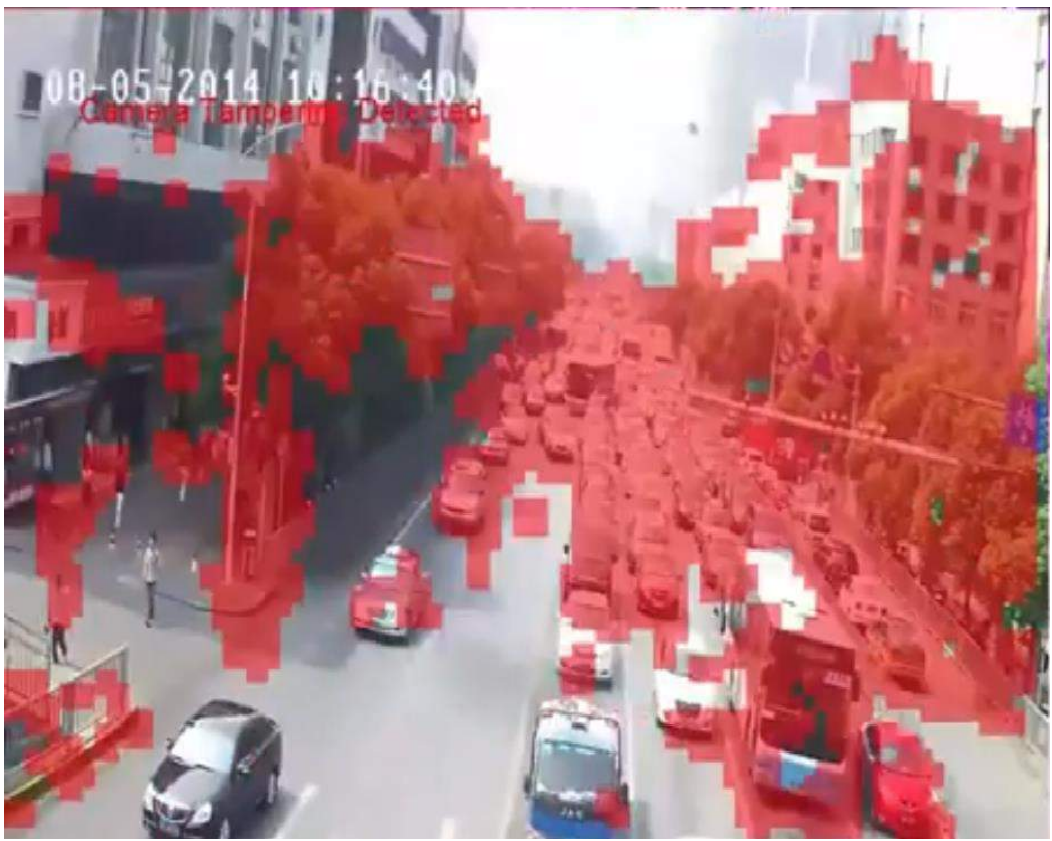

Fig.4.19 Camera Tampering – Camera redirection detection

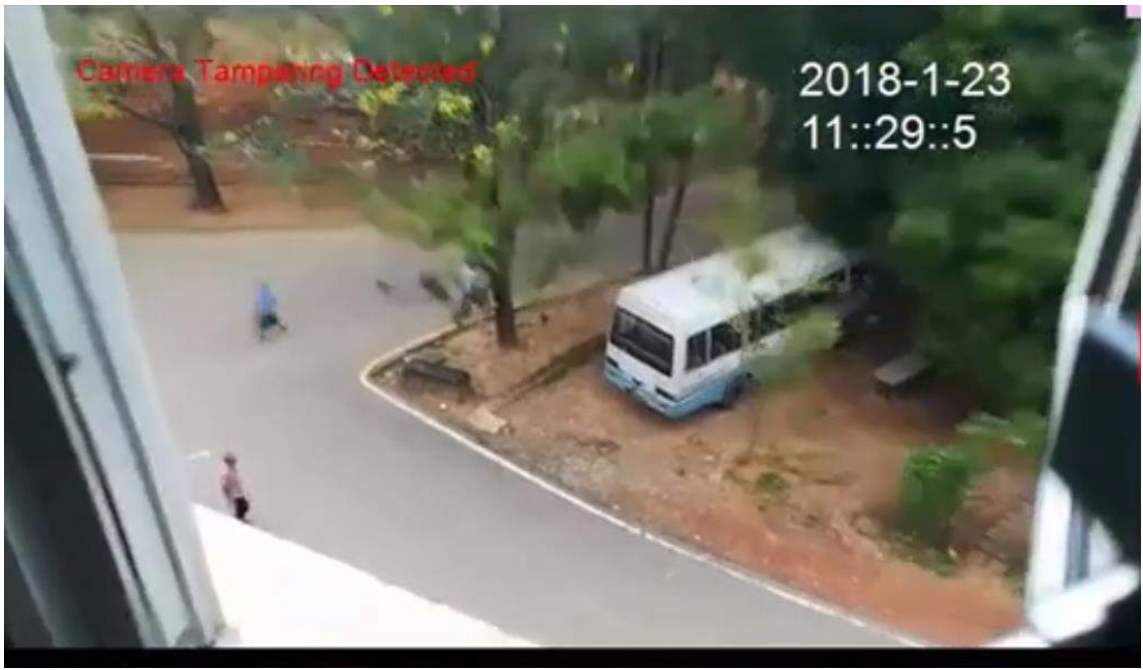

Fig4.20 Camera Tampering – Camera shaking detection

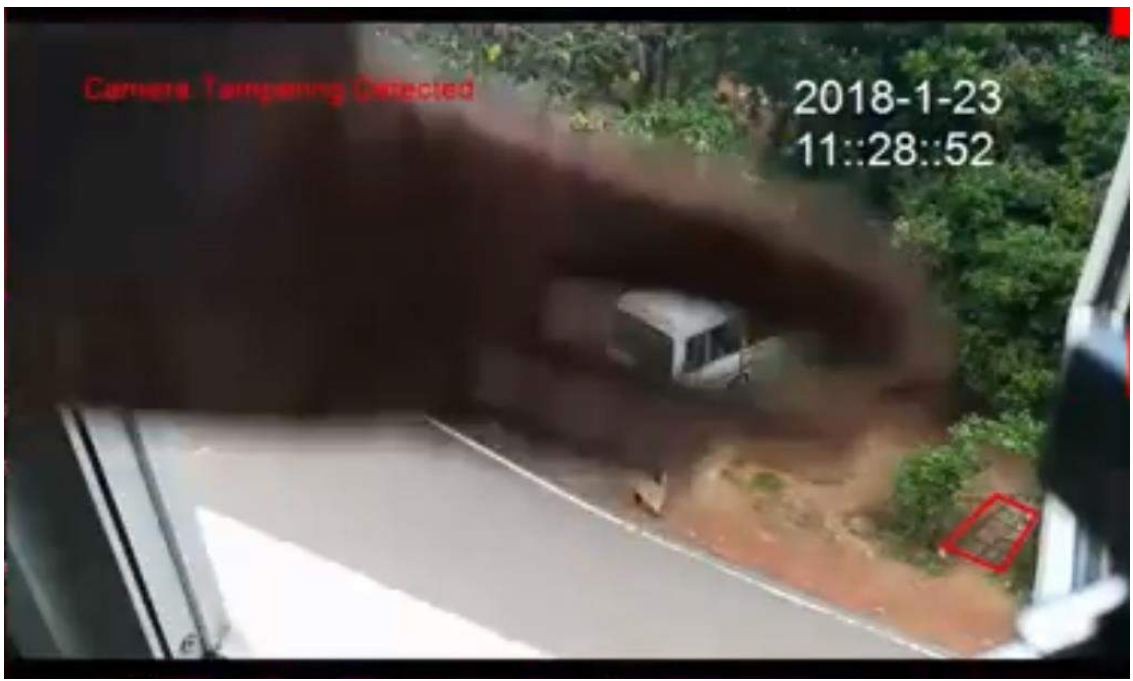

Fig.4.21 Camera Tampering – Camera block detection

**4.7 Summary**

From our quantitative and qualitative results, we can conclude that our software performs as expected in different applications we have tested as demonstrated in this chapter. The results obtained under forensic and surveillance architecture are presented separately. Further video synopsis, activity recognition, image enhancement and tampering detection results are given with emphasis on Sri Lankan context.

.

# Chapter 5

## DISCUSSION AND CONCLUSION

We developed an integrated solution for the problem of combining different techniques of video forensics and surveillance in an efficient manner. Although there are several implementations in this field of video forensic and surveillance as a collection of solutions, their performance is not efficient. Our goal was to come up with an integrated and efficient forensic video analytic software which is capable of processing CCTV feeds both in real time and for post processing.

Forensic Analytic Software can ease the work for Sri Lankan police officers and security force in detecting any forensic activities in real time and also help them to collect evidences for criminal activities. Features like video synopsis will save the time of the authorities in finding evidence for criminal activities and surveillance. Tampering detection enables the police to identify whether the produced evidence is being tampered or not. Activity recognition and anomaly detection enables real-time detection of activities which can cause crime and allows security officers to focus more on those scenes. Tracking multiple people in a video makes it easier to track the activities of a specified person. As the Sri Lankan police officers lack this technology, they are seeking help from lecturers in the Engineering Faculty of University of Moratuwa by sending their evidence video to extract some vital information. This Forensic Analytic Software will be an appropriate solution for their needs, as it is a user-friendly software they can enhance a region of interest (e.g. face or number plate) and collect the evidence that they require.

**5.1 Video Synopsis**

*5.1.1 Cluster Size*

Cluster size determines the number of tubes used to create synopsis video at each instance. It also thresholds the maximum number of tubes in a frame in synopsis video. When the cluster size is small, duration of synopsis video will be large while the synopsis video will be visually pleasing. When the cluster size is very large, synopsis video will be small while the synopsis video becomes visually unpleasant due to flickering and all the object frames of some of the tubes will not be placed in continuous order to avoid collision. Therefore, optimum cluster size should be chosen such that duration of synopsis video is minimized while the synopsis video is visually pleasant. Optimum cluster size also depends on the ratio of average object size to the size of region of interest occupied by moving objects. Since the number of objects that can be tracked by user in a frame also depends on the user, the proposed method allows user to set the cluster size.

Fig.5.1 shows that when cluster size is 5 most of the space are vacant, when the cluster size is 1000 region of interest occupied by moving object is completely packed and there is high chance that object frames of the tubes cannot be placed continuously due to different trajectories followed by different tubes. Cluster size of 20 produces optimal synopsis video for Cross Road video. In the below picture each object are labelled with different Object_IDs to track easily.

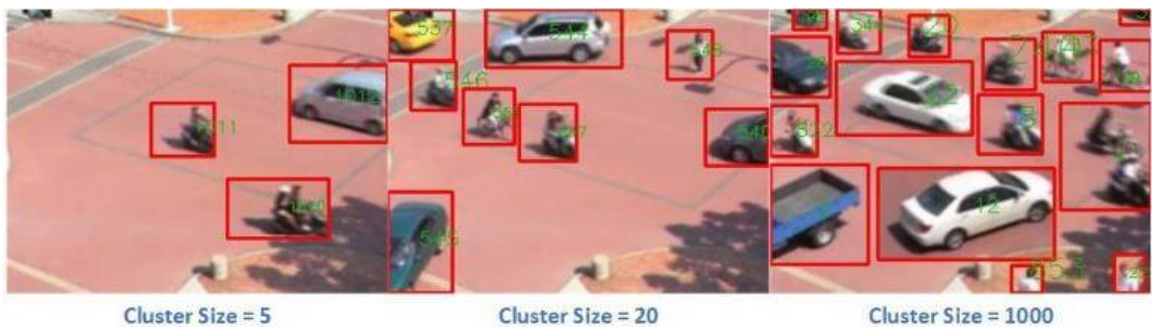

Fig.5.1 Synopsis video created for Cross Road video at different cluster size

### 5.1.2 *Failure Cases*

As the objects at different time instances are being placed in single frame in synopsis video and the background is updated asynchronously overtime in synopsis video, the following fault cases might arise.

- **Objects Overlap**

    As objects may become background overtime, moving objects may be placed over the objects which have become background. This may give fault impression to user that 2 objects are being overlapped. Figure 5.2 depicts an instance where black car has become background overtime when it is parked. The person who has walked through the area covered by car when it is not present in original video is depicted to walk over the black car in synopsis video.

- **Ghost Movement**

    As the number of frames used to train gaussian model for detection is limited to 100 frames in a 18FPS video. Any unusual slow movement like parking car will be suddenly depicted in synopsis video through background update. Fig.5.2 shows car parking scenario covered by background update.

- **Multiple Instances of the object**

    As an object may arrive through background update before its corresponding object tube, the same objects multiple instances may be seen in a single frame. Fig.5.2 shows corresponding instance.

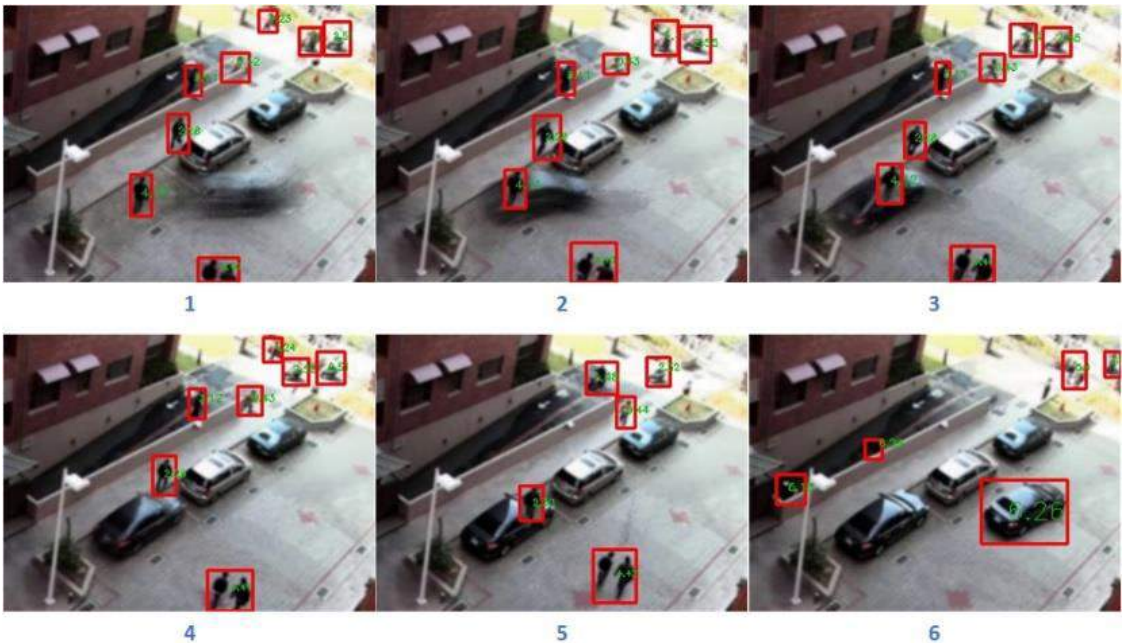

Fig.5.2: Fault case in Street Video

*5.1.3 Real-Time*

Despite of fault cases in situations which has less probability of occurrence, the proposed algorithm runs more than real-time in the above dataset. Therefore, the proposed algorithm can run in real-time even at higher density videos. This is very useful since CCTV cameras are being recorded 24 hours a day, because video synopsis should run in real-time, to synchronously summarize in real usage without any lag accumulation.

*5.1.4 Highways*

Proposed approach works well in highways at large cluster size since all the vehicles in a lane follows same trajectories and there is no issue of vehicle becoming background in normal situations as they are fast moving. Fig.5.3 shows synopsis video created at cluster size of 50 and 100 in M-30 and M-30-HD video respectively.

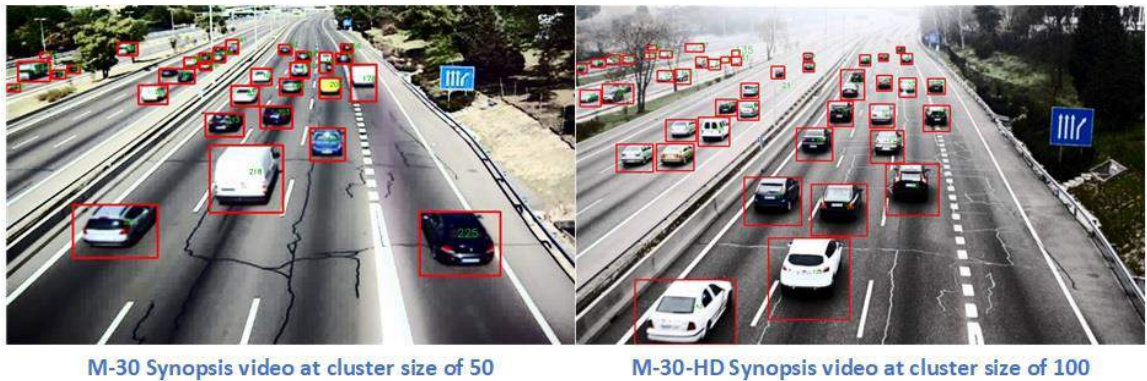

Fig.5.3 Synopsis video created with GRAM dataset

## 5.2 Forensic Architecture

In our project we have developed a unique forensic architecture for post processing. In this architecture, the user can use multiple forensic analysis tools on a video, a frame of interest in a video or an image to collect evidences easily.

In all enhancement tools the user can move sliders based on personal judgement until the image has been enhanced and the enhancement is only saved when the user clicks the "Apply" button. This allows the tool to be user-friendly since the user can recover the original image even if he makes erroneous slider movements.

The cropping tool also finalizes the cropping when the user clicks the "Apply" button. When the user feels one's cropping went wrong, he/she can recover the original image back by clicking the "Reset" button.

The user also can do detection of splice tampering and copy and move tampering on the same loaded image to confirm intra-frame tampering and whether the object has been inserted from the same image or if it is a foreign object.

We have carefully noticed the relationship between tools and we have efficiently managed variables and memory so that there will be no run time error when using multiple interrelated tools on a loaded image. This software has been tested by about ten users and the runtime errors detected by them have been corrected.

This architecture differs from other existing architectures because it contains most of the novel tools required to collect evidences under one platform.

**5.3 Surveillance Architecture**

Parallelism and real time implementation are the two vital performance measures to be considered in designing a surveillance application. Our novel architecture attempts to maximize the hardware resources and threads to implement the intended features in real-time.

It is directly deployable for a single camera application in real-time. We have enabled the communication between independent threads as well. Although the front end of the application (GUI) uses the Common Language Runtime (CLR) most of the inner layers are implemented on native C++, thus providing a high comfort to speed up the process and memory management of the unmanaged code.

If successfully optimized and implemented, this architecture is expandable, scalable and commercializeable. Further this architecture provides plug and play facility for more features and threads.

When tested in i7-7700K 4.2 GHz and NVIDIA GTX 1050 Ti 4 GB we were able to achieve a speed of 16 fps for an HD image which should be optimized further for more real-time application. Meanwhile the resource consumptions are as follows.

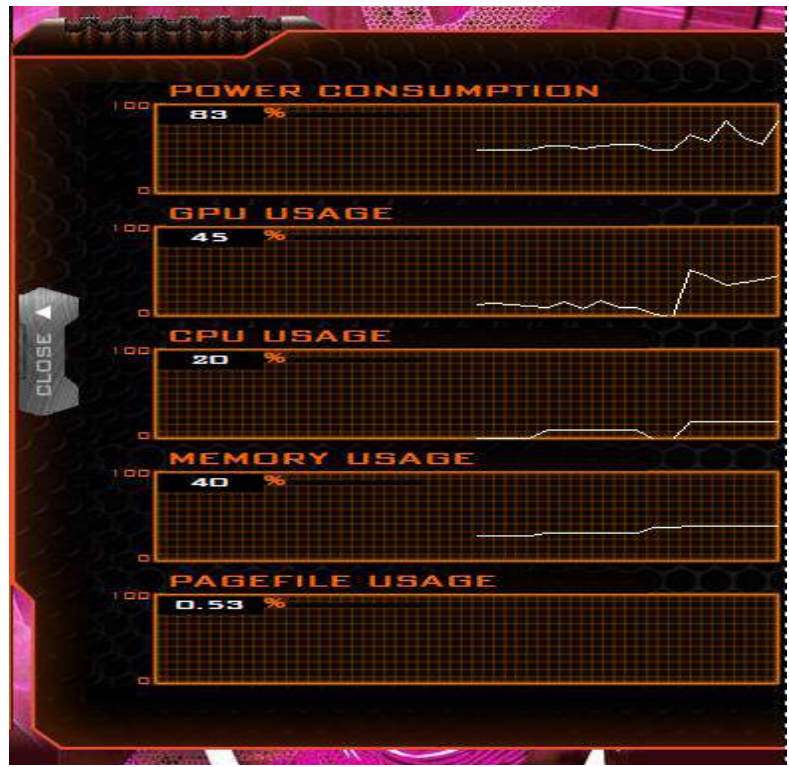

Fig.5.4 Resource consumption

Looking at the resource monitor, we can see clearly there are ample resources for expansion into additional threads and GPU usage. This will be considered in the future.

The GPU utilization peaks at about 50% indicating sufficient processing power for improvement. The current frame rate could be increased by adding multi-threading to the input video stream, eliminating the bottleneck prior to the synchronous blocks.

Activity recognition could be improved by using neural networks.

**5.4 Summary**

Summarizing this section, the results we gained indicate a considerable improvement in forensic video analytic software. Video Synopsis provides robust results. Object detection using YOLO provides accurate and fast results. Multiple object tracking provides online and real-time tracks. Activity recognition provides classifications between different activities. The effective integration of different modules into one software through a user-friendly GUI has been developed. These functionalities share some common processes which reduce the computational complexity and time in a considerable amount. Multi-threading too has an added advantage.

Based on our test on datasets available on the internet we have achieved satisfactory results, and for the practical implementation we have tested our work on self-recorded videos. Videos from the university premises too have been tested with. Anomaly detection performs well at Sri Lankan contest both in indoor and outdoor scenarios.

Overall, we have achieved a great result and it has opened a vast opportunity of practical implementation in the context of Sri Lanka as well as integrating with popular online platforms like Milestone. In this time, we acknowledge all those who helped us to achieve this.

# Chapter 6